\documentclass[aps,prb,floatfix,reprint,showpacs,superscriptaddress,longbibliography]{revtex4-2}
\usepackage{mathtools,amssymb,graphicx}
\usepackage{soul} 
\usepackage[dvipsnames,usenames]{color}
\usepackage[normalem]{ulem}
\usepackage{tabularx}
\usepackage{graphicx}
\usepackage{xcolor}
\usepackage{bbold, verbatim} 
\usepackage{float}
\usepackage{tabularx}
\usepackage{rotating}
\usepackage{braket}
\usepackage{multirow}
\usepackage{ulem}

\newcommand{\tvb}{\textsc{TurboRVB}} 
\newcommand{\tbg}{\textsc{TurboGenius}} 
\newcommand{\pyscf}{\textsc{PySCF}} 
\newcommand{\trexio}{\textsc{TREX-IO}} 

\begin{document} 
\title{Efficient calculation of unbiased atomic forces in {\emph{ab initio}} Variational Monte Carlo} 
\author{Kousuke Nakano}
\email{kousuke\_1123@icloud.com}
\affiliation{Center for Basic Research on Materials, National Institute for Materials Science (NIMS), Tsukuba, Ibaraki 305-0047, Japan}
\author{Michele Casula}
\affiliation{Institut de Min{\'e}ralogie, de Physique des Mat{\'e}riaux et de Cosmochimie (IMPMC), Sorbonne Universit{\'e}, CNRS UMR 7590, IRD UMR 206, MNHN, 4 Place Jussieu, 75252 Paris, France}
\author{Giacomo Tenti}
\affiliation{International School for Advanced Studies (SISSA), Via Bonomea 265, 34136, Trieste, Italy}

\date{\today}

\begin{abstract}
\emph{Ab initio} quantum Monte Carlo (QMC) is a state-of-the-art numerical approach for evaluating accurate expectation values of many-body wavefunctions. However, one of the major drawbacks that still hinders widespread QMC applications is the lack of an affordable scheme to compute unbiased atomic forces.
In this study, we propose an efficient method to obtain unbiased atomic forces and pressures in the Variational Monte Carlo (VMC) framework with the Jastrow-correlated Slater determinant ansatz or the Jastrow antisymmetrized geminal power ansatz, exploiting the gauge-invariant and locality properties of their geminal representation.  We demonstrate the effectiveness of our method for H$_2$ and Cl$_2$ molecules and for the cubic boron nitride crystal. Our framework has a better algorithmic scaling with the system size than the traditional finite-difference method, and, in practical applications, is as efficient as single-point VMC calculations. Thus, it paves the way to study dynamical properties of materials, such as phonons, and is beneficial for pursuing more reliable machine-learning interatomic potentials based on unbiased VMC forces.
\end{abstract}
\maketitle

\makeatletter
\def\Hline{
\noalign{\ifnum0=`}\fi\hrule \@height 1pt \futurelet
\reserved@a\@xhline}
\makeatother


\section{Introduction}
\label{sec:intro}

%
{\emph{Ab initio}} quantum Monte Carlo (QMC)~{\cite{2001FOU}} is a state-of-the-art numerical approach for evaluating the expectation values of many-body wavefunctions. It usually provides extremely accurate energies. To date, QMC has been successfully applied to various materials for which other electronic structure methods, such as the Density Functional Theory (DFT), lose predictive power. Examples are molecular crystals~{\cite{2018ZEN}}, two-dimensional materials~{\cite{2015MOS,2019FRA,2022NIK}}, superconductors~{\cite{2013CAS}}, and materials at extreme pressures~{\cite{2014CLAY,2016CLAY,2015DRU,2018MAZ,2023LOR}}.
Despite several successful applications done so far and the recent development of sophisticated QMC packages~{\cite{2009WAG, 2013SCE, 2020NEE, 2020PAU, 2023WHE}}, this technique is not as widely used as other established electronic structure methods. If compared with DFT~{\cite{2004MAR}}, one of the main QMC drawbacks is the lack of an efficient and affordable scheme to compute atomic forces consistent with the derivatives of the total energy with respect to atomic positions (a.k.a. unbiased atomic forces).
This problem is relevant in the construction of machine learning potentials (MLPs), which need large datasets, where energy and forces are computed with the method of choice. Recently, some QMC-driven MLPs have been reported~\cite{2021DIR,2022TIR1,2023HUA,2023NIU,2023CEP}, where the availability of unbiased forces and pressures has been a major concern. 
%

There are two main real-space QMC frameworks, the variational Monte Carlo (VMC) and the fixed-node diffusion Monte Carlo (FN-DMC) methods~{\cite{2001FOU}}. In this study, we focus on VMC because the forces computation within the FN-DMC framework is much more difficult and it is still a highly debated topic~{\cite{1986REY, 2000ASS, 2000FIL, 2005CHI, 2008BAD2, 2011ASS, 2014MOR, 2021VAN}}. Let ${\bf R}_{\alpha}$ be the atomic position of the nucleus $\alpha$. The atomic force acting on $\alpha$ is defined as the negative gradient of the energy with respect to ${\bf R}_{\alpha}$:
\begin{subequations}
\begin{align}
{\bf F}_{\alpha} = - \frac{dE}{d{\bf R}_{\alpha}} = &- \braket{\frac{\partial}{\partial {\bf R}_{\alpha}} E_{\rm L}} \label{eqn:hf} \\
&-2\braket{(E_{\rm L} - E) \frac{\partial \log \Psi_{\rm T}}{\partial {\bf R}_{\alpha}}}\label{eqn:pulay} \\
&-\sum_{i=1}^{N_p}{\frac{\partial E}{\partial p_{i}}} {\frac{d p_{i}}{d {\bf R}_{\alpha}}}, \label{eqn:add}
\end{align}
\end{subequations}
where $\Psi_{\rm T}$ is the variational wavefunction, $\braket{A}$ indicates the quantum average of the local operator $A$ over the VMC sampling of $|\Psi_{\rm T}|^2$, $E_{\rm L}$ is the so-called local energy ($E_{\rm L} \equiv \hat{H}\Psi_{\rm T}/\Psi_{\rm T}$), with $E \equiv \braket{E_L}$, and $\{ p_1, \cdots, p_{N_{p}}\}$ is the set of $N_{p}$ variational parameters included in the $\Psi_{\rm T}$ ansatz. Eqs.~(\ref{eqn:hf}), (\ref{eqn:pulay}), and (\ref{eqn:add}) are called the Hellmann--Feynman (HF), Pulay, and variational terms, respectively. 
One usually ignores Eq.~(\ref{eqn:add}) when evaluating atomic VMC forces, resulting in
\begin{equation}
{\bf F}_{\alpha}^{\rm{VMC}} = - \braket{\frac{\partial}{\partial {\bf R}_{\alpha}} E_{\rm L}} - 2 \braket{(E_{\rm L} - E) \frac{\partial \log \Psi_{\rm T}}{\partial {\bf R}_{\alpha}}}.
\label{eqn:vmc-force}
\end{equation}
%
The long-standing problem of obtaining a statistically meaningful ${\bf F}_{\alpha}^{\rm{VMC}}$ value with a finite variance and at the same cost as the VMC energy evaluation
has been solved by the zero-variance zero-bias principle~{\cite{2003ASS}} together with the space-warp transformation~{\cite{1989UMR} and reweighting techniques~{\cite{2000ASS, 2003ASS, 2008ATT, 2010SOR, 2016CLA, 2022NAK1}}. 
Hereafter, we will denote ${\bf F}_{\alpha}^{\rm{VMC}}$ as regular VMC force~(Eq.~\ref{eqn:vmc-force}). 

Neglecting Eq.~(\ref{eqn:add}) is justified only when the system is at its variational minimum for all parameters (i.e., $\partial E/\partial p_i=0$, $\forall i$) or when the variational parameters, which are {\it implicitly} dependent on the atomic positions, accidentally or by construction become position-independent (i.e., $d p_i/d {\bf R}_\alpha=0$, $\forall i$); otherwise, ${\bf F}_{\alpha}^{\rm{VMC}}$ can be biased. This bias is referred to as {\it self-consistency error}~{\cite{2021TII, 2022NAK1}}.
%


In this Letter, we propose a method to obtain unbiased atomic forces and pressures that does not increase the computational complexity of the VMC energy calculation, by supplementing the regular VMC force with a suitable variational term, computed by exploiting the gauge-invariant and locality properties of the antisymmetrized geminal power (AGP) ansatz~{\cite{2003CAS}}. For assessment, we demonstrate that the potential energy surfaces (PESs) of the H$_2$ and Cl$_2$ molecules, and the equation of state (EOS) of the cubic Boron Nitride (cBN) are consistent with the forces and pressure obtained by our proposed method.
%

\section{Illustrating the problem}
\label{sec:illustrating}
For the sake of clarity, we present the case of the PES of a dimer expressed as a function of the interatomic distance $R$, while the present discussion can be applied for any other system. Fig.~{\ref{fig:schematic-PES}} shows a schematic picture of several PESs. Let $E^{\rm exact}$(a) be the exact PES of the dimer. $E^{\rm exact}$ is the ultimate goal of any electronic structure calculation, but it is unknown except for nodeless ground states. The best possible PES within a given $\Psi_{\rm T}$ ansatz is $E^{\rm fullopt}$(b), yielded by a VMC calculation with the fully optimized $\Psi_{\rm T}$. This is achievable for rather small systems by optimization methods suitable for noisy data~{\cite{2005SOR, 2007UMR}}, but becomes impractical for larger ones. Therefore, a good compromise between accuracy and computational efficiency is $E_{\rm JSD}$(c), obtained by the Jastrow correlated Slater determinant (JSD) ansatz, with one-body molecular orbitals (MOs) computed by DFT for {\it each} interatomic distance $R$. The JSD is the most common VMC ansatz: only the Jastrow factor is optimized at the VMC level, while the DFT MOs are kept frozen in the Slater determinant (SD). However, in this case, the VMC force $F^{\rm{VMC}}$ is not consistent with the slope of $E_{\rm JSD}$, because the variational parameters included in the SD are not at their VMC minima. Instead, $F^{\rm{VMC}}$ corresponds to the slope of $E_{\rm JSD}^{\rm biased}$(d), where the DFT MOs obtained at $R = R'$ are used artificially for {\it all} $R$, such that $d p_i/d R=0$, $\forall i$. In this work, we propose an efficient method to obtain atomic forces and pressures that are unbiased, namely consistent with the slope of $E_{\rm JSD}$(c).

\begin{figure}[htbp]
  \centering
  \includegraphics[width=1.0\linewidth]{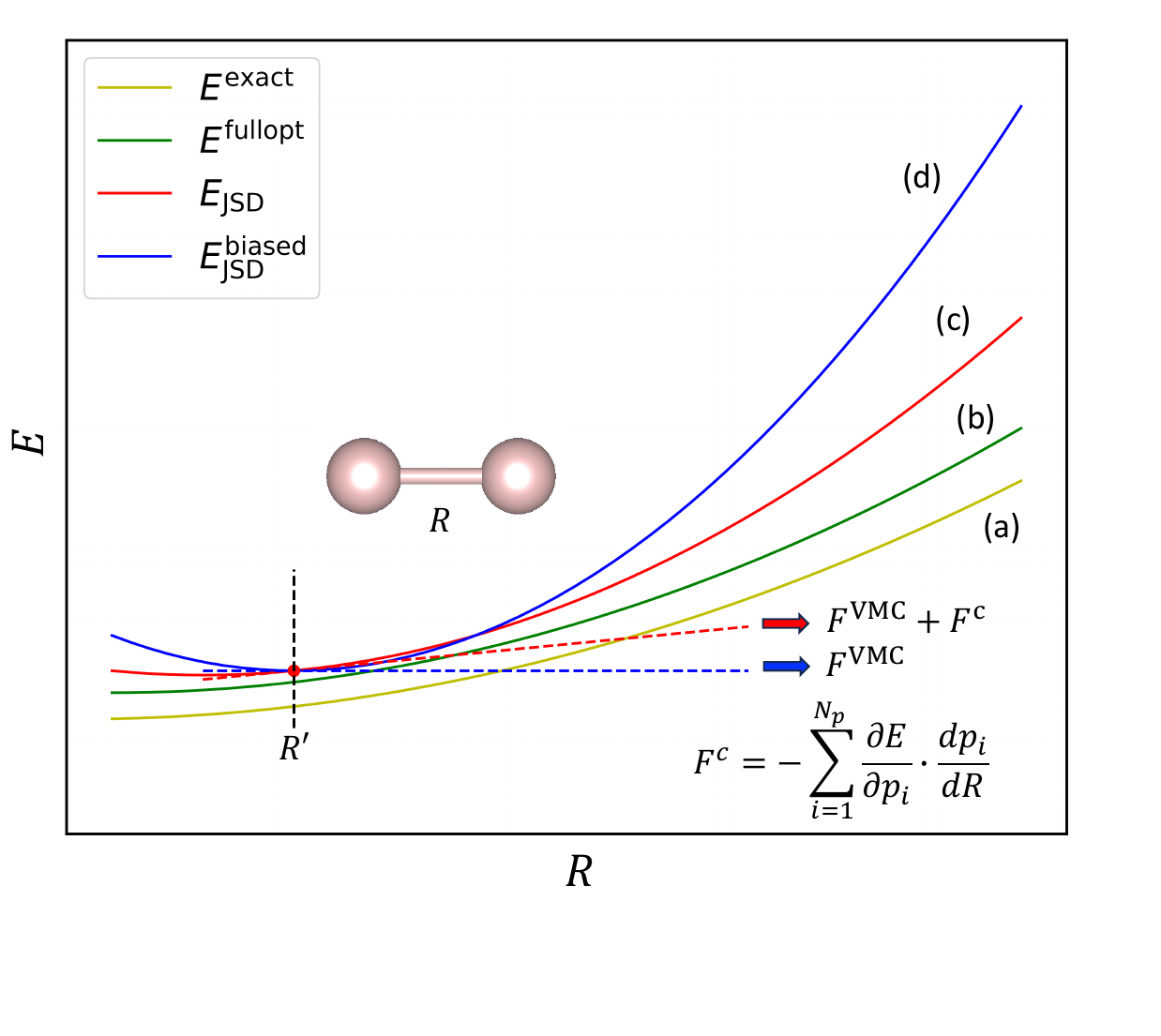}
  \caption{Schematic picture of PESs as a function of the dimer bond length $R$. (a) The exact PES, not accessible in practice. (b) The best possible PES obtained in the VMC framework by minimizing all variational parameters of $\Psi_{\rm T}$. (c) The PES obtained with optimized Jastrow factor and Slater MOs yielded by DFT at each point $R$, whose slope at $R^\prime$ is exactly given by the VMC force $F^{\rm{VMC}}$ supplemented by the variational term $F^c$, as proposed in this study. (d) The PES obtained with frozen DFT orbitals computed at $R'$, whose slope corresponds to $F^{\rm{VMC}}$ without the additional term $F^c$.}
  \label{fig:schematic-PES}
\end{figure}

\begin{figure*}[htbp]
  \centering
  \includegraphics[width=1.0\textwidth]{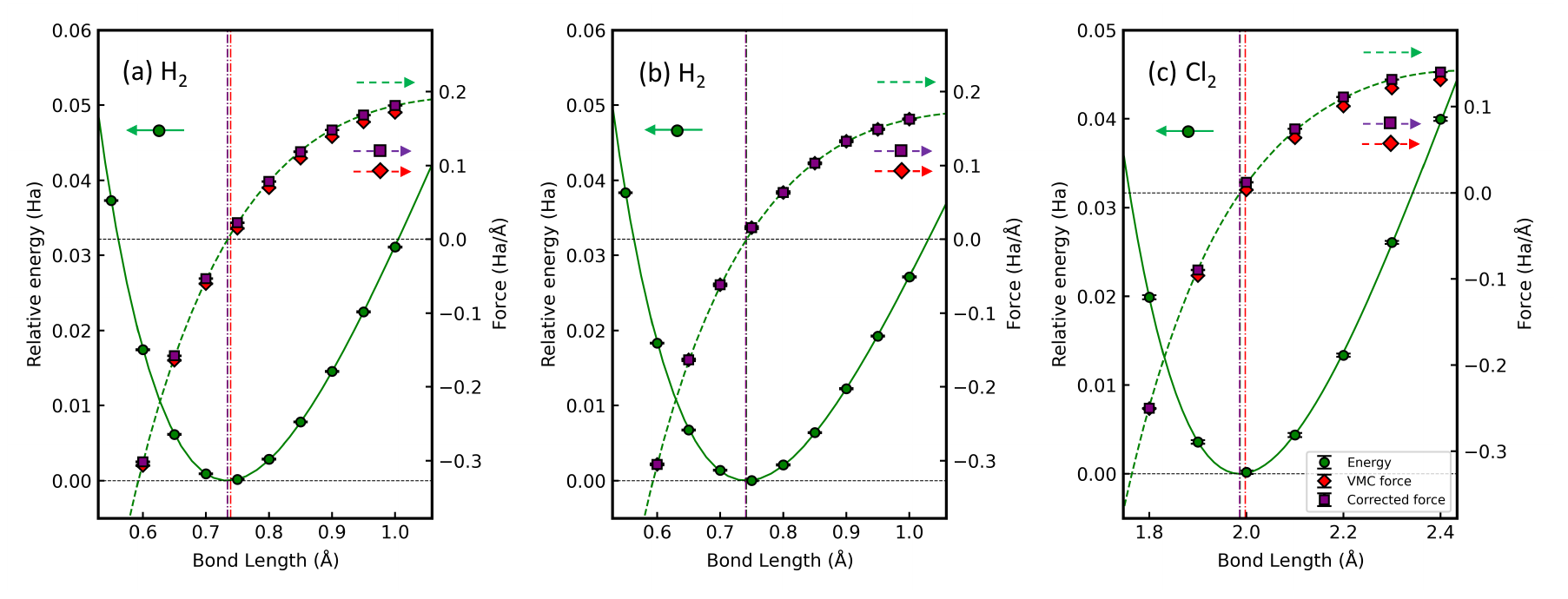}
  \caption{(a) and (b): H$_2$ PESs (solid green curves), their numerical derivatives (dashed green curves), regular VMC forces (red diamonds), and corrected forces (purple squares) obtained with (a) small [1s] and (b) large [4s2p1d] Jastrow basis sets. 
  The PESs and forces are computed from 0.30 \AA\ to 2.00 \AA\ with 18 equally spaced datapoints plus 5 additional datapoints (0.55 \AA, 0.65 \AA, 0.75 \AA, 0.85\AA, and 0.95 \AA). The vertical dashed lines represent equilibrium bond lengths obtained by fitting the PES (forces) with a polynomial of 11th (10th) order.
  (c): Cl$_2$ PES (solid green curve), its numerical derivative (dashed green curve), regular VMC forces (red diamonds), and corrected forces (purple squares). 
  The PES and forces are computed from 1.50 to 2.80 \AA\ with 14 equally spaced datapoints. The vertical dashed lines represent equilibrium bond lengths obtained by fitting the PES (forces) with a polynomial of 6th (5th) order for energies (forces).  
  In all panels, only the region in the vicinity of the equilibrium geometry is drawn.
  The plotted forces are $F_{x}$ acting on the left atom of each dimer, where the $x$ axis is aligned with the direction of the molecular bond.
  }
  \label{fig:h2-cl2-pes}
\end{figure*}

As long as the Jastrow factor is at its variational minimum, the contribution to the bias comes only from the SD part. This suggests that a straightforward solution for correcting the bias is to compute the variational term 
${\bf F}_{\alpha}^c \equiv -\sum_{i=1}^{N_p^{\rm SD}}{\cfrac{\partial E}{\partial p_{i}^{\rm SD}}} {\cfrac{d p_{i}^{\rm SD}}{d {\bf R}_{\alpha}}}$, where $p_i^{\rm SD}$ are SD variational parameters. In the following, we introduce a method to evaluate these terms by combining DFT with VMC gradients calculations.

\section{Method to obtain unbiased atomic forces}
\label{sec:method}
We begin by introducing the AGP representation~{\cite{2003CAS, 2020NAK2}} of the SD ansatz made of MOs. The general AGP ansatz for a system of $N_{\rm{e}}$ electrons is written as
%
$
\Psi_{\rm AGP} = \hat{\cal A} [
g({\bf x}_1,{\bf x}_2)
g({\bf x}_3,{\bf x}_4)
\ldots
g({\bf x}_{N_{\rm{e}}-1},{\bf x}_{N_{\rm{e}}})],
$
%
where $\hat{\cal A}$ is the antisymmetrization operator and $g$ is the so-called geminal function
$g({\bf x}_l,{\bf x}_m) = f({\bf r}_l, {\bf r}_m) (\ket{\uparrow\downarrow}-\ket{\downarrow\uparrow})/\sqrt{2}$.
The spatial part $f({\bf r}, \mathbf{r}^\prime)$ can be written in terms of MOs, such that
%
$f({\bf r}, \mathbf{r}^\prime) = \sum_{k}^M \Phi_k \,({\bf r}) \Phi_k({\bf r}^\prime)$,
%
where $\Phi_k(x)$ is the $k$-th MO expressed as $\Phi_k({\bf r}) = \sum_i^L c_{i, k} \psi_{i}({\bf r})$, $\psi_{i}({\bf r})$ is the $i$-th atomic orbital (AO), and, $c_{i, k}$ are the AO coefficients obtained by a DFT calculation. If the Hilbert space is restricted to the occupied states, i.e. $M=N_e/2$ for spin-unpolarized systems~{\footnote{The AGP ansatz has been generalized also to spin-polarized systems~{\cite{2003CAS}}.}}, the resultant AGP is equivalent to the SD ansatz. In other words, the SD ansatz can be treated as a special case of the more general AGP wavefunction. 
We assume that $\Psi_{\rm T}$ is real for the sake of conciseness; thus, the variational parameters are also real. However, our method can be readily generalized to complex $\Psi_{\rm T}$~{\cite{2017BEC}}.
In this work, the geminal function is constructed from the MOs obtained from a DFT calculation, and then converted to the AO representation, namely
%
$
f({\bf r}_l, {\bf r}_m) = \sum_{i,j}^{L,L} \lambda_{i, j} \psi_i({\bf r}_l) \psi_j({\bf r}_m),
$
%
with $\lambda_{i, j} = \sum_{k} c_{i,k} c_{j,k}$.
%
Thus, the variational term needed for correcting the self-consistency error reads
\begin{equation}\label{eq:fc}
{\bf F}_{\alpha}^{\rm c} = -\sum_{i,j}^{L,L}{\cfrac{\partial E}{\partial \lambda_{i,j}}} {\cfrac{d \lambda_{i,j}}{d {\bf R}_{\alpha}}},
\end{equation}
which is what one should compute to get unbiased atomic forces in the JSD ansatz, where $\lambda_{i,j}$ are directly obtained by DFT calculations. As discussed later, the geminal representation also allows one to compute unbiased forces and pressures beyond the JSD ansatz by optimizing a part of $\lambda_{i,j}$ in the JAGP ansatz at the VMC level.
%

The first factor in the terms summed in Eq.~\ref{eq:fc}, i.e. $\partial E/\partial \lambda_{i, j}$, used for optimizing $\Psi_{\rm T}$ and often dubbed as {\it generalized force}, can be efficiently computed by VMC~{\cite{2017BEC}}.
%
%
%
%
%
The second factor, the total derivative $d \lambda_{i,j}/d {{\bf R}_{\alpha}}$, can be numerically evaluated using the finite-difference method (FDM), i.e. $d \lambda_{i, j}/d {{\bf R}_{\alpha}} \sim \left(\lambda_{i, j}^{{{\bf R}_{\alpha}} + \Delta {{\bf R}_{\alpha}}} - \lambda_{i, j}^{{{{\bf R}_{\alpha}} - \Delta {{\bf R}_{\alpha}}}}\right)/2\Delta {{\bf R}_{\alpha}}$, or can be obtained by solving the coupled perturbed Hartree-Fock (CPHF) or Kohn-Sham (CPKS) equations~{\cite{2017JEN}}, or the linear response equations~{\cite{2018JUL}}. The second factor is $N$ times more time-consuming than the single-point DFT calculation, and this is regardless of the number of variational parameters. Indeed, to compute the correction terms for the geometry $\mathbf{R} \equiv \{{\bf R}_{1}, \ldots, {\bf R}_{N}\}$, one needs at least 3$N$-times HF/DFT calculations, where 3 is the number of Cartesian components. In this study, we employed the finite-difference approach because the {\it gauge-invariant} property of the AGP, inherited from its close relation with the reduced one-body density matrix~{\cite{1982OSV}}, allows one to construct a robust workflow to compute the second factor. Indeed, thanks to the gauge invariant property of the AGP, one does not suffer from (i) the global phase (or sign) indetermination of MOs, nor from (ii) their possible degeneracy.
As for (i), a global phase $\theta$ rotating the $k$-th MO ($\Phi_{k} \rightarrow e^{i\theta}\Phi_{k}$), which is reduced to a global sign $e^{i\theta} = \pm 1$ in case of a real $\Psi_{\rm T}$, does not affect the total energy, but prevents the  calculation of orbital derivatives, $d c_{i, k}/d {{\bf R}_{\alpha}}$, based on finite differences. Indeed, the global phase (or sign) is sometimes inconsistent between DFT outcomes with different atomic displacements. Instead, the sign flip is not problematic in the AGP representation, because the relation $\lambda_{i, j} = \sum_{k} c_{i,k} c_{j,k}$ 
implies that $\lambda_{i,j}$ is invariant under an MO sign change~{\footnote{$\lambda_{i,j}$ is invariant under a unitary transformation of MOs only if $\lambda_{i, j} = \sum_{k} c_{i,k}^* c_{j,k}$ in complex cases.}}.
As for (ii), 
when two (or more) MOs are degenerate, $c_{i, k}^{{{\bf R}_{\alpha}} + \Delta {{\bf R}_{\alpha}}}$ and $c_{i, k}^{{{\bf R}_{\alpha}} - \Delta {{\bf R}_{\alpha}}}$ might have very different values due to the presence of the other degenerate MOs.
%
%
%
Nevertheless, it is straightforward to show that an MOs degeneracy does not affect the uniqueness of $\lambda_{i,j}$, 
making $\lambda_{i,j}$ independent of the choice of the particular DFT implementation for degenerate MOs.
Thus, by exploiting the AO representation of the AGP wavefunction, one can always devise a well-defined method to compute $d\lambda_{i,j}/d {{\bf R}_{\alpha}}$, which will be superior to the calculation of $d c_{i, k}/d {{\bf R}_{\alpha}}$.

%

%
By combining the first and second factor in Eq.~\ref{eq:fc}, the variational term can be cast in a form suitable for a VMC estimate, as follows~{\footnote{Only its real part is taken in complex cases~{\cite{2017BEC}}}}:
%
\begin{eqnarray}\label{eq:fc-vmc-eval}
{\bf{F}}_{\alpha}^{\rm c} = -2\left\langle{E_{\rm L}(x) \sum_{i,j}^{L,L} \left[(O_{i,j}(x)-\bar{O}_{i,j}){\cfrac{d \lambda_{i,j}}{d {\bf R}_{\alpha}}}\right]}\right\rangle,
\end{eqnarray}
where we made apparent the dependence of the local operators on the total electronic coordinate $x$, sampled by VMC, to distinguish them from constant values.
In Eq.~\ref{eq:fc-vmc-eval}, $O_{i,j}(x) = \partial \ln \Psi_{T}(x)/\partial \lambda_{i,j}$, and $\bar{O}_{i,j} \sim \braket{O_{i,j}(x)}$. We remark that ${O}_{i,j}(x)$ can be efficiently computed in a VMC calculation using the adjoint algorithmic differentiation~{\cite{2010SOR}}, and the divergences of the generalized forces can be cured by reweighting methods~\cite{2008ATT, 2020PAT}.
It is extremely important that the variational term is evaluated 
in a covariance form of random variables to reduce its fluctuations~{\cite{2005SOR, 2005UMR}}. In addition, the expression in Eq.~\ref{eq:fc-vmc-eval} implies that if the variational wavefunction is an exact eigenstate of the Hamiltonian, ${\bf{F}}_{\alpha}^{\rm c}$ vanishes regardless of the VMC sample, because the local energy coincides with the corresponding eigenvalue $E$. Indeed, the zero-variance property holds in this expression, which is another way to recover the Hellmann-Feynman theorem. 

\section{Applications to H$_2$ and Cl$_2$ molecules}
\label{sec:applications-H2-Cl2}
We determine the interatomic force of the H$_2$ and Cl$_2$ molecules, taken as first examples to assess the accuracy of our method.
The ccECPs~{\cite{2017BEN,2018BEN,2018ANN,2019WAN}} accompanied with the uncontracted cc-pVDZ basis sets were employed for H$_2$ and Cl$_2$ molecules. For Cl$_2$, the He-core ccECP was employed. DFT-MOs were prepared by \pyscf\ v2.0.1~{\cite{2018SUN, 2020SUN}} with the LDA-PZ exchange-correlation functional~{\cite{1981PER}}, and then converted to the \tvb\ wavefunction format~{\cite{2020NAK2}} using the \tbg\ package~{\cite{2023NAK}} via \trexio\ files~{\cite{2023POS}}.
The inhomogeneous one-body, the two-body, and the three-body Jastrow factors~{\cite{2020NAK2}} were added to the SD with frozen DFT MOs and optimized using the linear method~{\cite{2005SOR, 2007UMR}} implemented in \tvb~{\cite{2020NAK2}}. The second factor, $d\lambda_{i,j}/d {{\bf R}_{\alpha}}$, was numerically evaluated using the displacements $\Delta R = \pm 0.001$ \AA\ along the molecular bond direction.

The simple H$_2$ molecule highlights the importance of removing the self-consistency error in the forces calculation by adding the variational force term to the regular VMC expression. 
In H$_2$, the JSD ansatz with DFT MOs is, in principle, exact if the Jastrow factor is converged in the basis set~{\footnote{See Ref.{~\cite{2020NAK2}} for the detail of the Jastrow factor implementation in the {\emph{ab initio}} QMC package, \tvb, used in this study.}}. Indeed, the wavefunction is nodeless, so the difficulty of finding the optimal variational state can be fully transferred to the Jastrow factor determination. Thus, H$_2$ allows one to study different situations, from a poor to a refined Jastrow factor. 
In this study, we examined 
a small [1s] and a large [4s2p1d] basis set expansion,
as a poor and refined Jastrow factor, respectively.
For the former, Fig.~{\ref{fig:h2-cl2-pes}}(a) shows 
that the DFT parameters are not optimal at the VMC level; thus, the self-consistency error is present. The equilibrium distance obtained from the PES (0.7344(2) \AA) and the one from regular VMC forces (0.7392(1) \AA) are reported in Tab.~{\ref{tab:PES-H2-Cl2}}. 
Fig.~{\ref{fig:h2-cl2-pes}} and Tab.~{\ref{tab:PES-H2-Cl2}} demonstrate that the self-consistency error is mitigated by the proposed force correction ${\bf{F}}^c$, which gives a bond distance 
of 0.7341(1) \AA, compatible with the one derived from the PES.
Fig.~{\ref{fig:h2-cl2-pes}}(b) shows that in the case of a refined Jastrow factor the self-consistency error is instead negligible, because the larger Jastrow expansion compensates for the DFT determinant, and all variational parameters are optimal.
Thus, the regular VMC force is already consistent with the derivative of the PES, and the corresponding force correction eventually vanishes, as reported in Tab.~{\ref{tab:PES-H2-Cl2}}. The H$_2$ example is illustrative of the capability of the variational term in Eq.~\ref{eq:fc} to correct the force bias due not only to the frozen DFT MOs, but also to an underconverged Jastrow factor.

Fig.~{\ref{fig:h2-cl2-pes}} (c) shows the PES of the Cl$_2$ molecule, as yielded by a [3s1p] Jastrow basis set. Tab.~{\ref{tab:PES-H2-Cl2}} reports the equilibrium geometries obtained from the PES, regular VMC force, and corrected force. The Figure and Table show that the self-consistency error is more significant for Cl$_2$ ($Z_{\rm eff} = 15$) than H$_2$ ($Z_{\rm eff} = 1$). This is consistent with the seminal work by Tiihonen {\it et al.}~{\cite{2021TII}}, reporting that the self-consistency error increases with the effective nuclear charge. Fig.~{\ref{fig:h2-cl2-pes}} and Tab.~{\ref{tab:PES-H2-Cl2}} illustrate that the proposed force correction works also for heavier molecules.


\begin{center}
\begin{table}[hbtp]
\caption{\label{tab:PES-H2-Cl2}
The equilibrium bond distances $r_{\rm eq}$ (\AA) of the H$_2$ and Cl$_2$ molecules obtained from the PESs, the regular VMC force, and the corrected force. The corresponding PESs are shown in Fig.~{\ref{fig:h2-cl2-pes}}.
}
\begin{tabular}{c|cc}
\Hline
Dimers & Source & $r_{{\rm eq}}$ (\AA) \\
\Hline
\multirow{4}{*}{H$_2$ (Jas.~[1s])} 
  & PES  &  0.7344(2)  \\
  & VMC force &  0.7392(1) \\
  & Corrected force &  0.7341(1) \\
  \cline{2-3}
  & Experiment  & 0.741~{\footnotemark[1]} \\
\Hline
\multirow{4}{*}{H$_2$ (Jas.[4s2p1d])}
  &  PES  &  0.7418(3)  \\
  & VMC force &  0.7408(6) \\
  & Corrected force &  0.7408(6) \\
  \cline{2-3}
  & Experiment  & 0.741~{\footnotemark[1]} \\
\Hline
\multirow{4}{*}{Cl$_2$ (Jas.[3s1p])}
  &  PES   &  1.987(1) \\
  &  VMC force &  1.9979(1) \\
  &  Corrected force &  1.9864(1) \\
  \cline{2-3}
  &  Experiment  & 1.987~{\footnotemark[1]} \\
\Hline
\end{tabular}
\footnotetext[1]{These values are taken from Ref.~\onlinecite{2013HUB}.
}
\end{table}
\end{center}

\section{Application to cubic Boron Nitride}
\label{sec:applications-to-cBN}
Not only atomic forces, but also pressures can be corrected in solids using the same method, just by replacing $d \lambda_{i, j}/d {{\bf R}_{\alpha}}$ with  $d \lambda_{i, j}/d V$. To demonstrate it, we computed the cBN EOS.
The ccECPs~{\cite{2017BEN,2018BEN,2018ANN,2019WAN}} with accompanying uncontracted cc-pVDZ basis sets were used for the cBN calculation. The linear dependency of the basis sets is solved at the DFT level by cutting basis set elements with exponents smaller than 0.20 a.u. This is crucial to suppress the statistical errors on atomic forces and pressures for periodic systems in QMC calculations~{\cite{2021NAK1}}. The 2$\times$2$\times$2 conventional supercell (256 valence electrons in the simulation cell) with $\mathbf{k}$ = $\Gamma$ was employed. DFT-MOs were prepared by the built-in DFT module implemented in \tvb~{\cite{2020NAK2}} with the LDA-PZ exchange-correlation functional~{\cite{1981PER}}. Then, the inhomogeneous one-body, the two-body, and the three-body Jastrow factors~{\cite{2020NAK2}} were added to the SD with frozen DFT molecular orbitals. [3s1p] Jastrow basis sets were employed for B and N atoms. The Jastrow factor was optimized using the linear method~{\cite{2005SOR, 2007UMR}} implemented in \tvb~{\cite{2020NAK2}} for each volume. The second factor, $d\lambda_{i,j}/d V$, was numerically evaluated using the built-in DFT module with volume variations $\Delta V = \pm 0.3$ \%.
Fig.~{\ref{fig:bn-eos}} shows the cBN EOS, its volume derivative, the regular VMC pressures, and the corrected pressures.
The obtained equilibrium lattice parameters and volumes are reported in Tab.~{\ref{tab:PES-cBN}}. It is apparent that 
the self-consistency error in pressure is $\sim 5$ GPa, constant over the whole volume range.
Our method gives corrections that bring the estimated pressures very close to the exact values for all volumes, as shown in Fig.~{\ref{fig:bn-eos}}. This result illustrates the possibility to successfully correct not only atomic forces but also pressures in large systems with the explicit evaluation of the variational pressure term. 

\begin{figure}[htbp]
  \centering
  \includegraphics[width=1.0\linewidth]{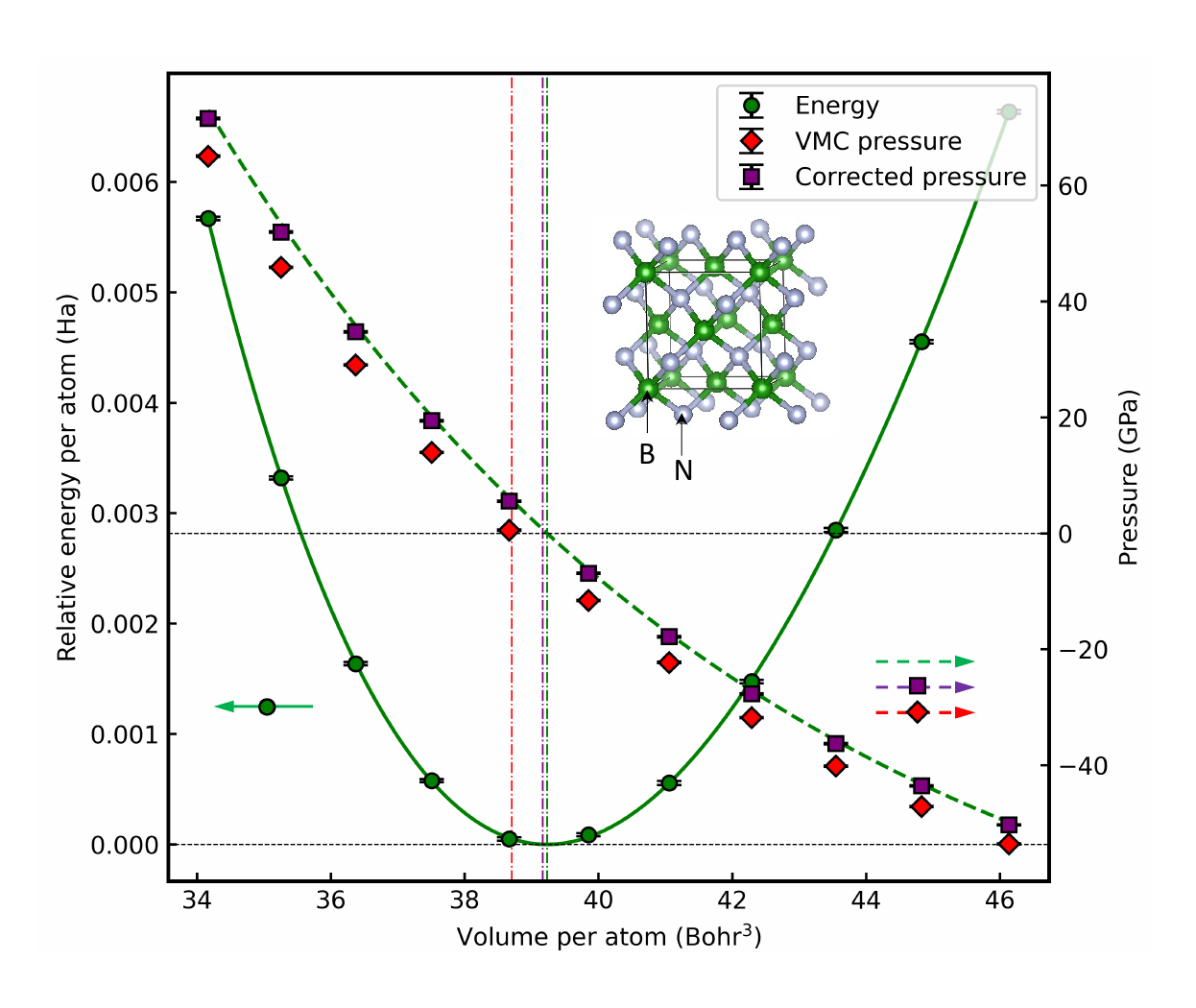}
  \caption{cBN EOS (solid green curve), its volume derivative (dashed green curve), the regular VMC pressure (red diamonds), and the corrected pressure (purple squares). The vertical dashed lines represent equilibrium volumes obtained by fitting the EOS and pressures with the Vinet forms~\cite{1987VIN}. 
  }
  \label{fig:bn-eos}
\end{figure}

\begin{center}
\begin{table}[hbtp]
\caption{\label{tab:PES-cBN}
Equilibrium lattice parameters and volumes per atom obtained by fitting the EOS, and from the regular VMC pressure and the corrected one. Zero point energy and temperature effects are not included.}
\begin{tabular}{c|cc}
\Hline
Source & Lattice (\AA) & Volume (Bohr$^3$) \\
\Hline
  EOS  &  3.5962(3) &  39.232(9) \\
  VMC pressure  &  3.5800(1) &  38.704(5) \\
  Corrected pressure  &  3.5943(2) &  39.169(5) \\
\hline
  Experiment  & 3.594~{\footnotemark[1]} & 39.160~{\footnotemark[1]} \\
\Hline
\end{tabular}
\footnotetext[1]{These values are taken from Ref.~\onlinecite{2013HUB}.}
\end{table}
\end{center}

\section{Discussion}
\label{sec:discussion}
We first compare our method with the FDM, which is the traditional way to obtain unbiased atomic forces in the VMC framework. 
The main drawback of the FDM is that it requires at least 3$N$ independent VMC runs to compute all 3$N$ force components, preventing its use in routine VMC calculations. Instead, our proposed method requires just a {\it single} VMC run to compute all 3$N$ regular VMC forces, together with all $O_{i,j}$ terms that appear in the expression for ${\bf{F}}_{\alpha}^{\rm c}$ of Eq.~\ref{eq:fc-vmc-eval}. This is thanks to the algorithmic differentiation~{\cite{2010SOR}}. As we mentioned before, the other terms in Eq.~\ref{eq:fc-vmc-eval}, namely $d \lambda_{i,j}/d {\bf R}_{\alpha}$, are computed by FDM using DFT as the driver, thus leading to DFT calculations $N$-times more time-consuming than a {\it single} DFT run. However, since the DFT cost is negligible compared to VMC, and it is mainly fast Fourier transform bound, with a favorable $O(N^2 \log N)$ scaling for a single run, the resulting algorithmic cost of our method is superior to the FDM evaluation of atomic VMC forces.

Next, we discuss the scaling of the variance of the variational term ${\bf{F}}_{\alpha}^{\rm c}$ with respect to $N$. The variance of the local energy $E_{\rm L}$ scales with $N_{\rm e}$~{\cite{2017BEC}}, while the variance of the logarithmic derivatives $O_{i,j}$ is $O(1)$~{\cite{2010SOR}}. Thus, the variance of ${\bf{F}}_{\alpha}^{\rm c}$ is bound by $O(L^2{N_{\rm e}})$, where the factor of $L^2$ comes from the double summation over the extended basis set elements~{\footnote{In the variance estimation, covariance effects are neglected. Nevertheless, in the actual calculations, the full statistical error was estimated by the Jackknife method~{\cite{2017BEC}}.}}.
The geminal representation needs the $L^2$ summation instead of the $LN_{\rm e}$ summation of the SD representation for the JSD ansatz.
However, at variance with the SD representation, the geminal allows one to 
exploit the {\it locality} of the $\lambda_{i,j}$ matrix. In other words, one can neglect $\left|d \lambda_{i,j}/d {\bf R}_{\alpha}\right|$ with small absolute values, obtained {\it deterministically} by DFT calculations.
For instance, the percentage of elements such that $\left|d \lambda_{i,j}/d {\bf R}_{\alpha}\right| / \max \left|d \lambda_{i,j}/d {\bf R}_{\alpha}\right| \le 0.01 \%$ is 38.5~\%, 45.0~\%, and 66.0~\% for 1 $\times$ 1 $\times$ 1 (8 atoms), 2 $\times$ 2 $\times$ 2 (64 atoms), and 3 $\times$ 3 $\times$ 3 (216 atoms) cBN supercells, respectively, demonstrating that the larger a system becomes, the more terms can be neglected thanks to the locality.
%
In this way, the summation in Eq.~\ref{eq:fc-vmc-eval} can be reduced from $L^2$ to $L$ terms, by lowering the size-scaling of the ${\bf F}_{\alpha}^{\rm c}$ variance.
Since $L$ and $N_{\rm e}$ are proportional to $N$, the scaling of the variance of our method with respect to $N$ is bound by $O(N^2)$ in the $N \rightarrow \infty$ limit, which is just 
$N$-times larger than the variance of the regular VMC force calculation, $O(N)$~{\cite{2010SOR}}. However, in our H$_2$, Cl$_2$, and cBN calculations, we got the same error bars on the regular VMC forces and on the corrected ones with the same statistics. This points to a very small prefactor $\varepsilon$ in the $O(N^2)$ variance term, such that in the total variance, ${\rm Var}({\bf F}_{\alpha}) = {\rm Var}({\bf F}_{\alpha}^{\rm{VMC}}) + {\rm Var} ({\bf F}_{\alpha}^{\rm{c}}) \approx O(N) + \varepsilon O(N^2)$,
the $O(N^2)$ contribution can be neglected for any affordable $N$ in VMC calculations.
Finally, we emphasize the extensibility of the geminal representation 
employed here, which allows one to readily generalize the method proposed in this work from the JSD to the more general JAGP ansatz. A practically way to go beyond the JSD ansatz for a large system is to optimize {\it only a subset} of the variational ${\lambda_{i,j}}$ parameters. The {\it partially} optimized ${\lambda_{i,j}}$ matrix will normally have a larger rank than the one corresponding to the SD wavefunction, therefore including AGP correlations. The subset of ${\lambda_{i,j}}$ is chosen again based on the AGP locality. Indeed, only the variational parameters ${\lambda_{i,j}}$ corresponding to atoms at a distance smaller than a reasonable cutoff can be optimized, while those with distance larger than the cutoff are kept fixed~{\cite{2017BEC}}. In this situation, only the fixed ${\lambda_{i,j}}$ must enter in Eq.~{\ref{eq:fc}}, thus correcting the force bias in the JAGP ansatz. In principle, our approach can also be extended to more general antisymmetric wavefunctions, with the only caveat that, in order to compute the $d p_i/d {\bf R}_\alpha$ derivatives, one has to consistently use the same auxiliary framework employed to initialize the antisymmetric part of the VMC wavefunction.

\section{Concluding remarks}
\label{sec:concluding-remarks}
In this work, we analyzed 
the bias seriously affecting the regular VMC expression ${\bf F}_{\alpha}^{\rm{VMC}}$.
We then proposed a method to efficiently and robustly compute the missing contribution, i.e. the variational term ${\bf F}_{\alpha}^{\rm{c}}$, to completely remove that bias
for a JSD ansatz with DFT one-body orbitals, the most common wavefunction in \emph{ab initio} VMC calculations, usually the best compromise between accuracy and computational cost. 
We demonstrated that the correction 
works very well for the systems that have been tested here, namely 
the equilibrium geometry of H$_2$ and Cl$_2$ molecules, and the EOS evaluation of the cubic Boron Nitride. Unbiased atomic forces within a JSD ansatz, which is in general much cheaper to optimize than more refined $\Psi_{\rm T}$s, will be particularly useful to generate VMC datasets for MLPs construction, which would otherwise be affected by the self-consistency error.
Thus, our approach has the potential to open up new horizons for VMC applications, also in the context of machine learning. 
Finally, the same scheme can be extended to more elaborated wavefunctions, once a suitable auxiliary method is used to generate their antisymmetric part.


\begin{acknowledgments}
K.N. is grateful for computational resources from the Numerical Materials Simulator at National Institute for Materials Science (NIMS).
The authors are grateful for computational resources of the supercomputer Fugaku provided by RIKEN through the HPCI System Research Projects (Project IDs: hp230030).
K.N. acknowledges financial support from Grant-in-Aid for Early Career Scientists (Grant No.~JP21K17752), from Grant-in-Aid for Scientific Research (Grant No.~JP21K03400), and from MEXT Leading Initiative for Excellent Young Researchers (Grant No.~JPMXS0320220025).
K.N. is grateful for fruitful discussion with Dr.~Atushi Togo (NIMS) about the illustration of the self-consistency problem.
The authors acknowledge valuable comments from Dr.~J\"{u}rg Hutter (UZH), Dr.~Stefano Battaglia (UZH) and Dr.~Emmanuel Giner (CNRS), Dr.~Saverio Moroni (SISSA), and Dr.~Claudia Filippi (UT).
M.C. acknowledges GENCI for granting access to French computational resources through the DARI application 906493.
The molecular and crystal structures were depicted using VESTA~{\cite{2011MOM}}.
This work is supported by the European Centre of Excellence in Exascale Computing TREX - Targeting Real Chemical Accuracy at the Exascale. This project has received funding from the European Union’s Horizon 2020 - Research and Innovation program - under grant agreement no. 952165.
The {\emph{ab initio}} QMC package used in this work, \tvb, is available from its GitHub repository [\url{https://github.com/sissaschool/turborvb}].
\end{acknowledgments}


\bibliographystyle{apsrev4-2}
\bibliography{./references.bib}

\begin{thebibliography}{65}%
\makeatletter
\providecommand \@ifxundefined [1]{%
 \@ifx{#1\undefined}
}%
\providecommand \@ifnum [1]{%
 \ifnum #1\expandafter \@firstoftwo
 \else \expandafter \@secondoftwo
 \fi
}%
\providecommand \@ifx [1]{%
 \ifx #1\expandafter \@firstoftwo
 \else \expandafter \@secondoftwo
 \fi
}%
\providecommand \natexlab [1]{#1}%
\providecommand \enquote  [1]{``#1''}%
\providecommand \bibnamefont  [1]{#1}%
\providecommand \bibfnamefont [1]{#1}%
\providecommand \citenamefont [1]{#1}%
\providecommand \href@noop [0]{\@secondoftwo}%
\providecommand \href [0]{\begingroup \@sanitize@url \@href}%
\providecommand \@href[1]{\@@startlink{#1}\@@href}%
\providecommand \@@href[1]{\endgroup#1\@@endlink}%
\providecommand \@sanitize@url [0]{\catcode `\\12\catcode `\$12\catcode `\&12\catcode `\#12\catcode `\^12\catcode `\_12\catcode `\%12\relax}%
\providecommand \@@startlink[1]{}%
\providecommand \@@endlink[0]{}%
\providecommand \url  [0]{\begingroup\@sanitize@url \@url }%
\providecommand \@url [1]{\endgroup\@href {#1}{\urlprefix }}%
\providecommand \urlprefix  [0]{URL }%
\providecommand \Eprint [0]{\href }%
\providecommand \doibase [0]{https://doi.org/}%
\providecommand \selectlanguage [0]{\@gobble}%
\providecommand \bibinfo  [0]{\@secondoftwo}%
\providecommand \bibfield  [0]{\@secondoftwo}%
\providecommand \translation [1]{[#1]}%
\providecommand \BibitemOpen [0]{}%
\providecommand \bibitemStop [0]{}%
\providecommand \bibitemNoStop [0]{.\EOS\space}%
\providecommand \EOS [0]{\spacefactor3000\relax}%
\providecommand \BibitemShut  [1]{\csname bibitem#1\endcsname}%
\let\auto@bib@innerbib\@empty
\bibitem [{\citenamefont {Foulkes}\ \emph {et~al.}(2001)\citenamefont {Foulkes}, \citenamefont {Mitas}, \citenamefont {Needs},\ and\ \citenamefont {Rajagopal}}]{2001FOU}%
  \BibitemOpen
  \bibfield  {author} {\bibinfo {author} {\bibfnamefont {W.~M.~C.}\ \bibnamefont {Foulkes}}, \bibinfo {author} {\bibfnamefont {L.}~\bibnamefont {Mitas}}, \bibinfo {author} {\bibfnamefont {R.~J.}\ \bibnamefont {Needs}},\ and\ \bibinfo {author} {\bibfnamefont {G.}~\bibnamefont {Rajagopal}},\ }\href {https://doi.org/10.1103/RevModPhys.73.33} {\bibfield  {journal} {\bibinfo  {journal} {Rev. Mod. Phys.}\ }\textbf {\bibinfo {volume} {73}},\ \bibinfo {pages} {33} (\bibinfo {year} {2001})}\BibitemShut {NoStop}%
\bibitem [{\citenamefont {Zen}\ \emph {et~al.}(2018)\citenamefont {Zen}, \citenamefont {Brandenburg}, \citenamefont {Klime{\v{s}}}, \citenamefont {Tkatchenko}, \citenamefont {Alf{\`e}},\ and\ \citenamefont {Michaelides}}]{2018ZEN}%
  \BibitemOpen
  \bibfield  {author} {\bibinfo {author} {\bibfnamefont {A.}~\bibnamefont {Zen}}, \bibinfo {author} {\bibfnamefont {J.~G.}\ \bibnamefont {Brandenburg}}, \bibinfo {author} {\bibfnamefont {J.}~\bibnamefont {Klime{\v{s}}}}, \bibinfo {author} {\bibfnamefont {A.}~\bibnamefont {Tkatchenko}}, \bibinfo {author} {\bibfnamefont {D.}~\bibnamefont {Alf{\`e}}},\ and\ \bibinfo {author} {\bibfnamefont {A.}~\bibnamefont {Michaelides}},\ }\href@noop {} {\bibfield  {journal} {\bibinfo  {journal} {Proc. Natl. Acad. Sci. U.S.A.}\ }\textbf {\bibinfo {volume} {115}},\ \bibinfo {pages} {1724} (\bibinfo {year} {2018})}\BibitemShut {NoStop}%
\bibitem [{\citenamefont {Mostaani}\ \emph {et~al.}(2015)\citenamefont {Mostaani}, \citenamefont {Drummond},\ and\ \citenamefont {Fal'ko}}]{2015MOS}%
  \BibitemOpen
  \bibfield  {author} {\bibinfo {author} {\bibfnamefont {E.}~\bibnamefont {Mostaani}}, \bibinfo {author} {\bibfnamefont {N.~D.}\ \bibnamefont {Drummond}},\ and\ \bibinfo {author} {\bibfnamefont {V.~I.}\ \bibnamefont {Fal'ko}},\ }\href {https://doi.org/10.1103/PhysRevLett.115.115501} {\bibfield  {journal} {\bibinfo  {journal} {Phys. Rev. Lett.}\ }\textbf {\bibinfo {volume} {115}},\ \bibinfo {pages} {115501} (\bibinfo {year} {2015})}\BibitemShut {NoStop}%
\bibitem [{\citenamefont {Frank}\ \emph {et~al.}(2019)\citenamefont {Frank}, \citenamefont {Derian}, \citenamefont {Tok\'ar}, \citenamefont {Mitas}, \citenamefont {Fabian},\ and\ \citenamefont {\ifmmode~\check{S}\else \v{S}\fi{}tich}}]{2019FRA}%
  \BibitemOpen
  \bibfield  {author} {\bibinfo {author} {\bibfnamefont {T.}~\bibnamefont {Frank}}, \bibinfo {author} {\bibfnamefont {R.}~\bibnamefont {Derian}}, \bibinfo {author} {\bibfnamefont {K.}~\bibnamefont {Tok\'ar}}, \bibinfo {author} {\bibfnamefont {L.}~\bibnamefont {Mitas}}, \bibinfo {author} {\bibfnamefont {J.}~\bibnamefont {Fabian}},\ and\ \bibinfo {author} {\bibfnamefont {I.}~\bibnamefont {\ifmmode~\check{S}\else \v{S}\fi{}tich}},\ }\href {https://doi.org/10.1103/PhysRevX.9.011018} {\bibfield  {journal} {\bibinfo  {journal} {Phys. Rev. X}\ }\textbf {\bibinfo {volume} {9}},\ \bibinfo {pages} {011018} (\bibinfo {year} {2019})}\BibitemShut {NoStop}%
\bibitem [{\citenamefont {Nikaido}\ \emph {et~al.}(2022)\citenamefont {Nikaido}, \citenamefont {Ichibha}, \citenamefont {Hongo}, \citenamefont {Reboredo}, \citenamefont {Kumar}, \citenamefont {Mahadevan}, \citenamefont {Maezono},\ and\ \citenamefont {Nakano}}]{2022NIK}%
  \BibitemOpen
  \bibfield  {author} {\bibinfo {author} {\bibfnamefont {Y.}~\bibnamefont {Nikaido}}, \bibinfo {author} {\bibfnamefont {T.}~\bibnamefont {Ichibha}}, \bibinfo {author} {\bibfnamefont {K.}~\bibnamefont {Hongo}}, \bibinfo {author} {\bibfnamefont {F.~A.}\ \bibnamefont {Reboredo}}, \bibinfo {author} {\bibfnamefont {K.~H.}\ \bibnamefont {Kumar}}, \bibinfo {author} {\bibfnamefont {P.}~\bibnamefont {Mahadevan}}, \bibinfo {author} {\bibfnamefont {R.}~\bibnamefont {Maezono}},\ and\ \bibinfo {author} {\bibfnamefont {K.}~\bibnamefont {Nakano}},\ }\href {https://doi.org/10.1021/acs.jpcc.1c10943} {\bibfield  {journal} {\bibinfo  {journal} {J. Phys. Chem. C}\ }\textbf {\bibinfo {volume} {126}},\ \bibinfo {pages} {6000} (\bibinfo {year} {2022})}\BibitemShut {NoStop}%
\bibitem [{\citenamefont {Casula}\ and\ \citenamefont {Sorella}(2013)}]{2013CAS}%
  \BibitemOpen
  \bibfield  {author} {\bibinfo {author} {\bibfnamefont {M.}~\bibnamefont {Casula}}\ and\ \bibinfo {author} {\bibfnamefont {S.}~\bibnamefont {Sorella}},\ }\href@noop {} {\bibfield  {journal} {\bibinfo  {journal} {Phys. Rev. B}\ }\textbf {\bibinfo {volume} {88}},\ \bibinfo {pages} {155125} (\bibinfo {year} {2013})}\BibitemShut {NoStop}%
\bibitem [{\citenamefont {Clay}\ \emph {et~al.}(2014)\citenamefont {Clay}, \citenamefont {Mcminis}, \citenamefont {McMahon}, \citenamefont {Pierleoni}, \citenamefont {Ceperley},\ and\ \citenamefont {Morales}}]{2014CLAY}%
  \BibitemOpen
  \bibfield  {author} {\bibinfo {author} {\bibfnamefont {R.~C.}\ \bibnamefont {Clay}}, \bibinfo {author} {\bibfnamefont {J.}~\bibnamefont {Mcminis}}, \bibinfo {author} {\bibfnamefont {J.~M.}\ \bibnamefont {McMahon}}, \bibinfo {author} {\bibfnamefont {C.}~\bibnamefont {Pierleoni}}, \bibinfo {author} {\bibfnamefont {D.~M.}\ \bibnamefont {Ceperley}},\ and\ \bibinfo {author} {\bibfnamefont {M.~A.}\ \bibnamefont {Morales}},\ }\href {https://doi.org/10.1103/PhysRevB.89.184106} {\bibfield  {journal} {\bibinfo  {journal} {Phys. Rev. B}\ }\textbf {\bibinfo {volume} {89}},\ \bibinfo {pages} {184106} (\bibinfo {year} {2014})}\BibitemShut {NoStop}%
\bibitem [{\citenamefont {Clay}\ \emph {et~al.}(2016)\citenamefont {Clay}, \citenamefont {Holzmann}, \citenamefont {Ceperley},\ and\ \citenamefont {Morales}}]{2016CLAY}%
  \BibitemOpen
  \bibfield  {author} {\bibinfo {author} {\bibfnamefont {R.~C.}\ \bibnamefont {Clay}}, \bibinfo {author} {\bibfnamefont {M.}~\bibnamefont {Holzmann}}, \bibinfo {author} {\bibfnamefont {D.~M.}\ \bibnamefont {Ceperley}},\ and\ \bibinfo {author} {\bibfnamefont {M.~A.}\ \bibnamefont {Morales}},\ }\href {https://doi.org/10.1103/PhysRevB.93.035121} {\bibfield  {journal} {\bibinfo  {journal} {Phys. Rev. B}\ }\textbf {\bibinfo {volume} {93}},\ \bibinfo {pages} {035121} (\bibinfo {year} {2016})}\BibitemShut {NoStop}%
\bibitem [{\citenamefont {Drummond}\ \emph {et~al.}(2015)\citenamefont {Drummond}, \citenamefont {Monserrat}, \citenamefont {Lloyd-Williams}, \citenamefont {R{\'\i}os}, \citenamefont {Pickard},\ and\ \citenamefont {Needs}}]{2015DRU}%
  \BibitemOpen
  \bibfield  {author} {\bibinfo {author} {\bibfnamefont {N.~D.}\ \bibnamefont {Drummond}}, \bibinfo {author} {\bibfnamefont {B.}~\bibnamefont {Monserrat}}, \bibinfo {author} {\bibfnamefont {J.~H.}\ \bibnamefont {Lloyd-Williams}}, \bibinfo {author} {\bibfnamefont {P.~L.}\ \bibnamefont {R{\'\i}os}}, \bibinfo {author} {\bibfnamefont {C.~J.}\ \bibnamefont {Pickard}},\ and\ \bibinfo {author} {\bibfnamefont {R.~J.}\ \bibnamefont {Needs}},\ }\href@noop {} {\bibfield  {journal} {\bibinfo  {journal} {Nat. Commun.}\ }\textbf {\bibinfo {volume} {6}},\ \bibinfo {pages} {1} (\bibinfo {year} {2015})}\BibitemShut {NoStop}%
\bibitem [{\citenamefont {Mazzola}\ \emph {et~al.}(2018)\citenamefont {Mazzola}, \citenamefont {Helled},\ and\ \citenamefont {Sorella}}]{2018MAZ}%
  \BibitemOpen
  \bibfield  {author} {\bibinfo {author} {\bibfnamefont {G.}~\bibnamefont {Mazzola}}, \bibinfo {author} {\bibfnamefont {R.}~\bibnamefont {Helled}},\ and\ \bibinfo {author} {\bibfnamefont {S.}~\bibnamefont {Sorella}},\ }\href@noop {} {\bibfield  {journal} {\bibinfo  {journal} {Phys. Rev. Lett.}\ }\textbf {\bibinfo {volume} {120}},\ \bibinfo {pages} {025701} (\bibinfo {year} {2018})}\BibitemShut {NoStop}%
\bibitem [{\citenamefont {Monacelli}\ \emph {et~al.}(2023)\citenamefont {Monacelli}, \citenamefont {Casula}, \citenamefont {Nakano}, \citenamefont {Sorella},\ and\ \citenamefont {Mauri}}]{2023LOR}%
  \BibitemOpen
  \bibfield  {author} {\bibinfo {author} {\bibfnamefont {L.}~\bibnamefont {Monacelli}}, \bibinfo {author} {\bibfnamefont {M.}~\bibnamefont {Casula}}, \bibinfo {author} {\bibfnamefont {K.}~\bibnamefont {Nakano}}, \bibinfo {author} {\bibfnamefont {S.}~\bibnamefont {Sorella}},\ and\ \bibinfo {author} {\bibfnamefont {F.}~\bibnamefont {Mauri}},\ }\href {https://doi.org/10.1038/s41567-023-01960-5} {\bibfield  {journal} {\bibinfo  {journal} {Nat. Phys.}\ }\textbf {\bibinfo {volume} {19}},\ \bibinfo {pages} {845} (\bibinfo {year} {2023})}\BibitemShut {NoStop}%
\bibitem [{\citenamefont {Wagner}\ \emph {et~al.}(2009)\citenamefont {Wagner}, \citenamefont {Bajdich},\ and\ \citenamefont {Mitas}}]{2009WAG}%
  \BibitemOpen
  \bibfield  {author} {\bibinfo {author} {\bibfnamefont {L.~K.}\ \bibnamefont {Wagner}}, \bibinfo {author} {\bibfnamefont {M.}~\bibnamefont {Bajdich}},\ and\ \bibinfo {author} {\bibfnamefont {L.}~\bibnamefont {Mitas}},\ }\href@noop {} {\bibfield  {journal} {\bibinfo  {journal} {J. Comput. Phys.}\ }\textbf {\bibinfo {volume} {228}},\ \bibinfo {pages} {3390} (\bibinfo {year} {2009})}\BibitemShut {NoStop}%
\bibitem [{\citenamefont {Scemama}\ \emph {et~al.}(2013)\citenamefont {Scemama}, \citenamefont {Caffarel}, \citenamefont {Oseret},\ and\ \citenamefont {Jalby}}]{2013SCE}%
  \BibitemOpen
  \bibfield  {author} {\bibinfo {author} {\bibfnamefont {A.}~\bibnamefont {Scemama}}, \bibinfo {author} {\bibfnamefont {M.}~\bibnamefont {Caffarel}}, \bibinfo {author} {\bibfnamefont {E.}~\bibnamefont {Oseret}},\ and\ \bibinfo {author} {\bibfnamefont {W.}~\bibnamefont {Jalby}},\ }\href@noop {} {\bibfield  {journal} {\bibinfo  {journal} {J. Comput. Chem.}\ }\textbf {\bibinfo {volume} {34}},\ \bibinfo {pages} {938} (\bibinfo {year} {2013})}\BibitemShut {NoStop}%
\bibitem [{\citenamefont {Needs}\ \emph {et~al.}(2020)\citenamefont {Needs}, \citenamefont {Towler}, \citenamefont {Drummond}, \citenamefont {Lopez~Rios},\ and\ \citenamefont {Trail}}]{2020NEE}%
  \BibitemOpen
  \bibfield  {author} {\bibinfo {author} {\bibfnamefont {R.}~\bibnamefont {Needs}}, \bibinfo {author} {\bibfnamefont {M.}~\bibnamefont {Towler}}, \bibinfo {author} {\bibfnamefont {N.}~\bibnamefont {Drummond}}, \bibinfo {author} {\bibfnamefont {P.}~\bibnamefont {Lopez~Rios}},\ and\ \bibinfo {author} {\bibfnamefont {J.}~\bibnamefont {Trail}},\ }\href@noop {} {\bibfield  {journal} {\bibinfo  {journal} {J. Chem. Phys.}\ }\textbf {\bibinfo {volume} {152}},\ \bibinfo {pages} {154106} (\bibinfo {year} {2020})}\BibitemShut {NoStop}%
\bibitem [{\citenamefont {Kent}\ \emph {et~al.}(2020)\citenamefont {Kent}, \citenamefont {Annaberdiyev}, \citenamefont {Benali}, \citenamefont {Bennett}, \citenamefont {Landinez~Borda}, \citenamefont {Doak}, \citenamefont {Hao}, \citenamefont {Jordan}, \citenamefont {Krogel}, \citenamefont {Kylänpää}, \citenamefont {Lee}, \citenamefont {Luo}, \citenamefont {Malone}, \citenamefont {Melton}, \citenamefont {Mitas}, \citenamefont {Morales}, \citenamefont {Neuscamman}, \citenamefont {Reboredo}, \citenamefont {Rubenstein}, \citenamefont {Saritas}, \citenamefont {Upadhyay}, \citenamefont {Wang}, \citenamefont {Zhang},\ and\ \citenamefont {Zhao}}]{2020PAU}%
  \BibitemOpen
  \bibfield  {author} {\bibinfo {author} {\bibfnamefont {P.~R.~C.}\ \bibnamefont {Kent}}, \bibinfo {author} {\bibfnamefont {A.}~\bibnamefont {Annaberdiyev}}, \bibinfo {author} {\bibfnamefont {A.}~\bibnamefont {Benali}}, \bibinfo {author} {\bibfnamefont {M.~C.}\ \bibnamefont {Bennett}}, \bibinfo {author} {\bibfnamefont {E.~J.}\ \bibnamefont {Landinez~Borda}}, \bibinfo {author} {\bibfnamefont {P.}~\bibnamefont {Doak}}, \bibinfo {author} {\bibfnamefont {H.}~\bibnamefont {Hao}}, \bibinfo {author} {\bibfnamefont {K.~D.}\ \bibnamefont {Jordan}}, \bibinfo {author} {\bibfnamefont {J.~T.}\ \bibnamefont {Krogel}}, \bibinfo {author} {\bibfnamefont {I.}~\bibnamefont {Kylänpää}}, \bibinfo {author} {\bibfnamefont {J.}~\bibnamefont {Lee}}, \bibinfo {author} {\bibfnamefont {Y.}~\bibnamefont {Luo}}, \bibinfo {author} {\bibfnamefont {F.~D.}\ \bibnamefont {Malone}}, \bibinfo {author} {\bibfnamefont {C.~A.}\ \bibnamefont {Melton}}, \bibinfo {author} {\bibfnamefont {L.}~\bibnamefont {Mitas}}, \bibinfo {author} {\bibfnamefont
  {M.~A.}\ \bibnamefont {Morales}}, \bibinfo {author} {\bibfnamefont {E.}~\bibnamefont {Neuscamman}}, \bibinfo {author} {\bibfnamefont {F.~A.}\ \bibnamefont {Reboredo}}, \bibinfo {author} {\bibfnamefont {B.}~\bibnamefont {Rubenstein}}, \bibinfo {author} {\bibfnamefont {K.}~\bibnamefont {Saritas}}, \bibinfo {author} {\bibfnamefont {S.}~\bibnamefont {Upadhyay}}, \bibinfo {author} {\bibfnamefont {G.}~\bibnamefont {Wang}}, \bibinfo {author} {\bibfnamefont {S.}~\bibnamefont {Zhang}},\ and\ \bibinfo {author} {\bibfnamefont {L.}~\bibnamefont {Zhao}},\ }\href@noop {} {\bibfield  {journal} {\bibinfo  {journal} {J. Chem. Phys.}\ }\textbf {\bibinfo {volume} {152}} (\bibinfo {year} {2020})}\BibitemShut {NoStop}%
\bibitem [{\citenamefont {Wheeler}\ \emph {et~al.}(2023)\citenamefont {Wheeler}, \citenamefont {Pathak}, \citenamefont {Kleiner}, \citenamefont {Yuan}, \citenamefont {Rodrigues}, \citenamefont {Lorsung}, \citenamefont {Krongchon}, \citenamefont {Chang}, \citenamefont {Zhou}, \citenamefont {Busemeyer}, \citenamefont {Williams}, \citenamefont {Muñoz}, \citenamefont {Chow},\ and\ \citenamefont {Wagner}}]{2023WHE}%
  \BibitemOpen
  \bibfield  {author} {\bibinfo {author} {\bibfnamefont {W.~A.}\ \bibnamefont {Wheeler}}, \bibinfo {author} {\bibfnamefont {S.}~\bibnamefont {Pathak}}, \bibinfo {author} {\bibfnamefont {K.~G.}\ \bibnamefont {Kleiner}}, \bibinfo {author} {\bibfnamefont {S.}~\bibnamefont {Yuan}}, \bibinfo {author} {\bibfnamefont {J.~N.~B.}\ \bibnamefont {Rodrigues}}, \bibinfo {author} {\bibfnamefont {C.}~\bibnamefont {Lorsung}}, \bibinfo {author} {\bibfnamefont {K.}~\bibnamefont {Krongchon}}, \bibinfo {author} {\bibfnamefont {Y.}~\bibnamefont {Chang}}, \bibinfo {author} {\bibfnamefont {Y.}~\bibnamefont {Zhou}}, \bibinfo {author} {\bibfnamefont {B.}~\bibnamefont {Busemeyer}}, \bibinfo {author} {\bibfnamefont {K.~T.}\ \bibnamefont {Williams}}, \bibinfo {author} {\bibfnamefont {A.}~\bibnamefont {Muñoz}}, \bibinfo {author} {\bibfnamefont {C.~Y.}\ \bibnamefont {Chow}},\ and\ \bibinfo {author} {\bibfnamefont {L.~K.}\ \bibnamefont {Wagner}},\ }\href {https://doi.org/10.1063/5.0139024} {\bibfield  {journal} {\bibinfo  {journal} {J.
  Chem. Phys.}\ }\textbf {\bibinfo {volume} {158}},\ \bibinfo {pages} {114801} (\bibinfo {year} {2023})}\BibitemShut {NoStop}%
\bibitem [{\citenamefont {Martin}(2004)}]{2004MAR}%
  \BibitemOpen
  \bibfield  {author} {\bibinfo {author} {\bibfnamefont {R.~M.}\ \bibnamefont {Martin}},\ }\href@noop {} {\emph {\bibinfo {title} {{Electronic structure : basic theory and practical methods}}}}\ (\bibinfo  {publisher} {Cambridge University Press},\ \bibinfo {year} {2004})\BibitemShut {NoStop}%
\bibitem [{\citenamefont {DiRisio}\ \emph {et~al.}(2021)\citenamefont {DiRisio}, \citenamefont {Lu},\ and\ \citenamefont {McCoy}}]{2021DIR}%
  \BibitemOpen
  \bibfield  {author} {\bibinfo {author} {\bibfnamefont {R.~J.}\ \bibnamefont {DiRisio}}, \bibinfo {author} {\bibfnamefont {F.}~\bibnamefont {Lu}},\ and\ \bibinfo {author} {\bibfnamefont {A.~B.}\ \bibnamefont {McCoy}},\ }\href {https://doi.org/10.1021/acs.jpca.1c03709} {\bibfield  {journal} {\bibinfo  {journal} {J. Phys. Chem. A}\ }\textbf {\bibinfo {volume} {125}},\ \bibinfo {pages} {5849} (\bibinfo {year} {2021})}\BibitemShut {NoStop}%
\bibitem [{\citenamefont {Tirelli}\ \emph {et~al.}(2022)\citenamefont {Tirelli}, \citenamefont {Tenti}, \citenamefont {Nakano},\ and\ \citenamefont {Sorella}}]{2022TIR1}%
  \BibitemOpen
  \bibfield  {author} {\bibinfo {author} {\bibfnamefont {A.}~\bibnamefont {Tirelli}}, \bibinfo {author} {\bibfnamefont {G.}~\bibnamefont {Tenti}}, \bibinfo {author} {\bibfnamefont {K.}~\bibnamefont {Nakano}},\ and\ \bibinfo {author} {\bibfnamefont {S.}~\bibnamefont {Sorella}},\ }\href {https://doi.org/10.1103/PhysRevB.106.L041105} {\bibfield  {journal} {\bibinfo  {journal} {Phys. Rev. B}\ }\textbf {\bibinfo {volume} {106}},\ \bibinfo {pages} {L041105} (\bibinfo {year} {2022})}\BibitemShut {NoStop}%
\bibitem [{\citenamefont {Huang}\ and\ \citenamefont {Rubenstein}(2023)}]{2023HUA}%
  \BibitemOpen
  \bibfield  {author} {\bibinfo {author} {\bibfnamefont {C.}~\bibnamefont {Huang}}\ and\ \bibinfo {author} {\bibfnamefont {B.~M.}\ \bibnamefont {Rubenstein}},\ }\href {https://doi.org/10.1021/acs.jpca.2c05904} {\bibfield  {journal} {\bibinfo  {journal} {J. Phys. Chem. A}\ }\textbf {\bibinfo {volume} {127}},\ \bibinfo {pages} {339} (\bibinfo {year} {2023})}\BibitemShut {NoStop}%
\bibitem [{\citenamefont {Niu}\ \emph {et~al.}(2023)\citenamefont {Niu}, \citenamefont {Yang}, \citenamefont {Jensen}, \citenamefont {Holzmann}, \citenamefont {Pierleoni},\ and\ \citenamefont {Ceperley}}]{2023NIU}%
  \BibitemOpen
  \bibfield  {author} {\bibinfo {author} {\bibfnamefont {H.}~\bibnamefont {Niu}}, \bibinfo {author} {\bibfnamefont {Y.}~\bibnamefont {Yang}}, \bibinfo {author} {\bibfnamefont {S.}~\bibnamefont {Jensen}}, \bibinfo {author} {\bibfnamefont {M.}~\bibnamefont {Holzmann}}, \bibinfo {author} {\bibfnamefont {C.}~\bibnamefont {Pierleoni}},\ and\ \bibinfo {author} {\bibfnamefont {D.~M.}\ \bibnamefont {Ceperley}},\ }\href {https://doi.org/10.1103/PhysRevLett.130.076102} {\bibfield  {journal} {\bibinfo  {journal} {Phys. Rev. Lett.}\ }\textbf {\bibinfo {volume} {130}},\ \bibinfo {pages} {076102} (\bibinfo {year} {2023})}\BibitemShut {NoStop}%
\bibitem [{\citenamefont {Ceperley}\ \emph {et~al.}(2024)\citenamefont {Ceperley}, \citenamefont {Jensen}, \citenamefont {Yang}, \citenamefont {Niu}, \citenamefont {Pierleoni},\ and\ \citenamefont {Holzmann}}]{2023CEP}%
  \BibitemOpen
  \bibfield  {author} {\bibinfo {author} {\bibfnamefont {D.}~\bibnamefont {Ceperley}}, \bibinfo {author} {\bibfnamefont {S.}~\bibnamefont {Jensen}}, \bibinfo {author} {\bibfnamefont {Y.~P.}\ \bibnamefont {Yang}}, \bibinfo {author} {\bibfnamefont {H.}~\bibnamefont {Niu}}, \bibinfo {author} {\bibfnamefont {C.}~\bibnamefont {Pierleoni}},\ and\ \bibinfo {author} {\bibfnamefont {M.}~\bibnamefont {Holzmann}},\ }\href@noop {} {\bibfield  {journal} {\bibinfo  {journal} {Electron. Struct.}\ }\textbf {\bibinfo {volume} {in press}} (\bibinfo {year} {2024})}\BibitemShut {NoStop}%
\bibitem [{\citenamefont {Reynolds}\ \emph {et~al.}(1986)\citenamefont {Reynolds}, \citenamefont {Barnett}, \citenamefont {Hammond}, \citenamefont {Grimes},\ and\ \citenamefont {Lester~Jr}}]{1986REY}%
  \BibitemOpen
  \bibfield  {author} {\bibinfo {author} {\bibfnamefont {P.}~\bibnamefont {Reynolds}}, \bibinfo {author} {\bibfnamefont {R.}~\bibnamefont {Barnett}}, \bibinfo {author} {\bibfnamefont {B.}~\bibnamefont {Hammond}}, \bibinfo {author} {\bibfnamefont {R.}~\bibnamefont {Grimes}},\ and\ \bibinfo {author} {\bibfnamefont {W.}~\bibnamefont {Lester~Jr}},\ }\href@noop {} {\bibfield  {journal} {\bibinfo  {journal} {Int. J. Quantum Chem.}\ }\textbf {\bibinfo {volume} {29}},\ \bibinfo {pages} {589} (\bibinfo {year} {1986})}\BibitemShut {NoStop}%
\bibitem [{\citenamefont {Assaraf}\ and\ \citenamefont {Caffarel}(2000)}]{2000ASS}%
  \BibitemOpen
  \bibfield  {author} {\bibinfo {author} {\bibfnamefont {R.}~\bibnamefont {Assaraf}}\ and\ \bibinfo {author} {\bibfnamefont {M.}~\bibnamefont {Caffarel}},\ }\href@noop {} {\bibfield  {journal} {\bibinfo  {journal} {J. Chem. Phys.}\ }\textbf {\bibinfo {volume} {113}},\ \bibinfo {pages} {4028} (\bibinfo {year} {2000})}\BibitemShut {NoStop}%
\bibitem [{\citenamefont {Filippi}\ and\ \citenamefont {Umrigar}(2000)}]{2000FIL}%
  \BibitemOpen
  \bibfield  {author} {\bibinfo {author} {\bibfnamefont {C.}~\bibnamefont {Filippi}}\ and\ \bibinfo {author} {\bibfnamefont {C.~J.}\ \bibnamefont {Umrigar}},\ }\href {https://doi.org/10.1103/PhysRevB.61.R16291} {\bibfield  {journal} {\bibinfo  {journal} {Phys. Rev. B}\ }\textbf {\bibinfo {volume} {61}},\ \bibinfo {pages} {R16291} (\bibinfo {year} {2000})}\BibitemShut {NoStop}%
\bibitem [{\citenamefont {Chiesa}\ \emph {et~al.}(2005)\citenamefont {Chiesa}, \citenamefont {Ceperley},\ and\ \citenamefont {Zhang}}]{2005CHI}%
  \BibitemOpen
  \bibfield  {author} {\bibinfo {author} {\bibfnamefont {S.}~\bibnamefont {Chiesa}}, \bibinfo {author} {\bibfnamefont {D.~M.}\ \bibnamefont {Ceperley}},\ and\ \bibinfo {author} {\bibfnamefont {S.}~\bibnamefont {Zhang}},\ }\href {https://doi.org/10.1103/PhysRevLett.94.036404} {\bibfield  {journal} {\bibinfo  {journal} {Phys. Rev. Lett.}\ }\textbf {\bibinfo {volume} {94}},\ \bibinfo {pages} {036404} (\bibinfo {year} {2005})}\BibitemShut {NoStop}%
\bibitem [{\citenamefont {Badinski}\ and\ \citenamefont {Needs}(2008)}]{2008BAD2}%
  \BibitemOpen
  \bibfield  {author} {\bibinfo {author} {\bibfnamefont {A.}~\bibnamefont {Badinski}}\ and\ \bibinfo {author} {\bibfnamefont {R.~J.}\ \bibnamefont {Needs}},\ }\href {https://doi.org/10.1103/PhysRevB.78.035134} {\bibfield  {journal} {\bibinfo  {journal} {Phys. Rev. B}\ }\textbf {\bibinfo {volume} {78}},\ \bibinfo {pages} {035134} (\bibinfo {year} {2008})}\BibitemShut {NoStop}%
\bibitem [{\citenamefont {Assaraf}\ \emph {et~al.}(2011)\citenamefont {Assaraf}, \citenamefont {Caffarel},\ and\ \citenamefont {Kollias}}]{2011ASS}%
  \BibitemOpen
  \bibfield  {author} {\bibinfo {author} {\bibfnamefont {R.}~\bibnamefont {Assaraf}}, \bibinfo {author} {\bibfnamefont {M.}~\bibnamefont {Caffarel}},\ and\ \bibinfo {author} {\bibfnamefont {A.~C.}\ \bibnamefont {Kollias}},\ }\href {https://doi.org/10.1103/PhysRevLett.106.150601} {\bibfield  {journal} {\bibinfo  {journal} {Phys. Rev. Lett.}\ }\textbf {\bibinfo {volume} {106}},\ \bibinfo {pages} {150601} (\bibinfo {year} {2011})}\BibitemShut {NoStop}%
\bibitem [{\citenamefont {Moroni}\ \emph {et~al.}(2014)\citenamefont {Moroni}, \citenamefont {Saccani},\ and\ \citenamefont {Filippi}}]{2014MOR}%
  \BibitemOpen
  \bibfield  {author} {\bibinfo {author} {\bibfnamefont {S.}~\bibnamefont {Moroni}}, \bibinfo {author} {\bibfnamefont {S.}~\bibnamefont {Saccani}},\ and\ \bibinfo {author} {\bibfnamefont {C.}~\bibnamefont {Filippi}},\ }\href {https://doi.org/10.1021/ct500780r} {\bibfield  {journal} {\bibinfo  {journal} {J. Chem. Theory Comput.}\ }\textbf {\bibinfo {volume} {10}},\ \bibinfo {pages} {4823} (\bibinfo {year} {2014})}\BibitemShut {NoStop}%
\bibitem [{\citenamefont {Van~Rhijn}\ \emph {et~al.}(2021)\citenamefont {Van~Rhijn}, \citenamefont {Filippi}, \citenamefont {De~Palo},\ and\ \citenamefont {Moroni}}]{2021VAN}%
  \BibitemOpen
  \bibfield  {author} {\bibinfo {author} {\bibfnamefont {J.}~\bibnamefont {Van~Rhijn}}, \bibinfo {author} {\bibfnamefont {C.}~\bibnamefont {Filippi}}, \bibinfo {author} {\bibfnamefont {S.}~\bibnamefont {De~Palo}},\ and\ \bibinfo {author} {\bibfnamefont {S.}~\bibnamefont {Moroni}},\ }\href {https://doi.org/10.1021/acs.jctc.1c00496} {\bibfield  {journal} {\bibinfo  {journal} {J. Chem. Theory Comput.}\ }\textbf {\bibinfo {volume} {18}},\ \bibinfo {pages} {118} (\bibinfo {year} {2021})}\BibitemShut {NoStop}%
\bibitem [{\citenamefont {Assaraf}\ and\ \citenamefont {Caffarel}(2003)}]{2003ASS}%
  \BibitemOpen
  \bibfield  {author} {\bibinfo {author} {\bibfnamefont {R.}~\bibnamefont {Assaraf}}\ and\ \bibinfo {author} {\bibfnamefont {M.}~\bibnamefont {Caffarel}},\ }\href {https://doi.org/10.1063/1.1621615} {\bibfield  {journal} {\bibinfo  {journal} {J. Chem. Phys.}\ }\textbf {\bibinfo {volume} {119}},\ \bibinfo {pages} {10536} (\bibinfo {year} {2003})}\BibitemShut {NoStop}%
\bibitem [{\citenamefont {Umrigar}(1989)}]{1989UMR}%
  \BibitemOpen
  \bibfield  {author} {\bibinfo {author} {\bibfnamefont {C.~J.}\ \bibnamefont {Umrigar}},\ }\href@noop {} {\bibfield  {journal} {\bibinfo  {journal} {Int. J. Quantum Chem.}\ }\textbf {\bibinfo {volume} {36}},\ \bibinfo {pages} {217} (\bibinfo {year} {1989})}\BibitemShut {NoStop}%
\bibitem [{\citenamefont {Attaccalite}\ and\ \citenamefont {Sorella}(2008)}]{2008ATT}%
  \BibitemOpen
  \bibfield  {author} {\bibinfo {author} {\bibfnamefont {C.}~\bibnamefont {Attaccalite}}\ and\ \bibinfo {author} {\bibfnamefont {S.}~\bibnamefont {Sorella}},\ }\href@noop {} {\bibfield  {journal} {\bibinfo  {journal} {Phys. Rev. Lett.}\ }\textbf {\bibinfo {volume} {100}},\ \bibinfo {pages} {114501} (\bibinfo {year} {2008})}\BibitemShut {NoStop}%
\bibitem [{\citenamefont {Sorella}\ and\ \citenamefont {Capriotti}(2010)}]{2010SOR}%
  \BibitemOpen
  \bibfield  {author} {\bibinfo {author} {\bibfnamefont {S.}~\bibnamefont {Sorella}}\ and\ \bibinfo {author} {\bibfnamefont {L.}~\bibnamefont {Capriotti}},\ }\href@noop {} {\bibfield  {journal} {\bibinfo  {journal} {J. Chem. Phys.}\ }\textbf {\bibinfo {volume} {133}},\ \bibinfo {pages} {234111} (\bibinfo {year} {2010})}\BibitemShut {NoStop}%
\bibitem [{\citenamefont {Filippi}\ \emph {et~al.}(2016)\citenamefont {Filippi}, \citenamefont {Assaraf},\ and\ \citenamefont {Moroni}}]{2016CLA}%
  \BibitemOpen
  \bibfield  {author} {\bibinfo {author} {\bibfnamefont {C.}~\bibnamefont {Filippi}}, \bibinfo {author} {\bibfnamefont {R.}~\bibnamefont {Assaraf}},\ and\ \bibinfo {author} {\bibfnamefont {S.}~\bibnamefont {Moroni}},\ }\href {https://doi.org/10.1063/1.4948778} {\bibfield  {journal} {\bibinfo  {journal} {J. Chem. Phys.}\ }\textbf {\bibinfo {volume} {144}},\ \bibinfo {pages} {194105} (\bibinfo {year} {2016})}\BibitemShut {NoStop}%
\bibitem [{\citenamefont {Nakano}\ \emph {et~al.}(2022)\citenamefont {Nakano}, \citenamefont {Raghav},\ and\ \citenamefont {Sorella}}]{2022NAK1}%
  \BibitemOpen
  \bibfield  {author} {\bibinfo {author} {\bibfnamefont {K.}~\bibnamefont {Nakano}}, \bibinfo {author} {\bibfnamefont {A.}~\bibnamefont {Raghav}},\ and\ \bibinfo {author} {\bibfnamefont {S.}~\bibnamefont {Sorella}},\ }\href {https://doi.org/10.1063/5.0076302} {\bibfield  {journal} {\bibinfo  {journal} {J. Chem. Phys.}\ }\textbf {\bibinfo {volume} {156}},\ \bibinfo {pages} {034101} (\bibinfo {year} {2022})}\BibitemShut {NoStop}%
\bibitem [{\citenamefont {Tiihonen}\ \emph {et~al.}(2021)\citenamefont {Tiihonen}, \citenamefont {Clay~III},\ and\ \citenamefont {Krogel}}]{2021TII}%
  \BibitemOpen
  \bibfield  {author} {\bibinfo {author} {\bibfnamefont {J.}~\bibnamefont {Tiihonen}}, \bibinfo {author} {\bibfnamefont {R.~C.}\ \bibnamefont {Clay~III}},\ and\ \bibinfo {author} {\bibfnamefont {J.~T.}\ \bibnamefont {Krogel}},\ }\href {https://doi.org/10.1063/5.0052266} {\bibfield  {journal} {\bibinfo  {journal} {J. Chem. Phys.}\ }\textbf {\bibinfo {volume} {154}},\ \bibinfo {pages} {204111} (\bibinfo {year} {2021})}\BibitemShut {NoStop}%
\bibitem [{\citenamefont {Casula}\ and\ \citenamefont {Sorella}(2003)}]{2003CAS}%
  \BibitemOpen
  \bibfield  {author} {\bibinfo {author} {\bibfnamefont {M.}~\bibnamefont {Casula}}\ and\ \bibinfo {author} {\bibfnamefont {S.}~\bibnamefont {Sorella}},\ }\href {https://doi.org/10.1063/1.1604379} {\bibfield  {journal} {\bibinfo  {journal} {J. Chem. Phys.}\ }\textbf {\bibinfo {volume} {119}},\ \bibinfo {pages} {6500} (\bibinfo {year} {2003})}\BibitemShut {NoStop}%
\bibitem [{\citenamefont {Sorella}(2005)}]{2005SOR}%
  \BibitemOpen
  \bibfield  {author} {\bibinfo {author} {\bibfnamefont {S.}~\bibnamefont {Sorella}},\ }\href@noop {} {\bibfield  {journal} {\bibinfo  {journal} {Phys. Rev. B}\ }\textbf {\bibinfo {volume} {71}},\ \bibinfo {pages} {241103(R)} (\bibinfo {year} {2005})}\BibitemShut {NoStop}%
\bibitem [{\citenamefont {Umrigar}\ \emph {et~al.}(2007)\citenamefont {Umrigar}, \citenamefont {Toulouse}, \citenamefont {Filippi}, \citenamefont {Sorella},\ and\ \citenamefont {Hennig}}]{2007UMR}%
  \BibitemOpen
  \bibfield  {author} {\bibinfo {author} {\bibfnamefont {C.~J.}\ \bibnamefont {Umrigar}}, \bibinfo {author} {\bibfnamefont {J.}~\bibnamefont {Toulouse}}, \bibinfo {author} {\bibfnamefont {C.}~\bibnamefont {Filippi}}, \bibinfo {author} {\bibfnamefont {S.}~\bibnamefont {Sorella}},\ and\ \bibinfo {author} {\bibfnamefont {R.~G.}\ \bibnamefont {Hennig}},\ }\href {https://doi.org/10.1103/PhysRevLett.98.110201} {\bibfield  {journal} {\bibinfo  {journal} {Phys. Rev. Lett.}\ }\textbf {\bibinfo {volume} {98}},\ \bibinfo {pages} {110201} (\bibinfo {year} {2007})}\BibitemShut {NoStop}%
\bibitem [{\citenamefont {Nakano}\ \emph {et~al.}(2020)\citenamefont {Nakano}, \citenamefont {Attaccalite}, \citenamefont {Barborini}, \citenamefont {Capriotti}, \citenamefont {Casula}, \citenamefont {Coccia}, \citenamefont {Dagrada}, \citenamefont {Genovese}, \citenamefont {Luo}, \citenamefont {Mazzola}, \citenamefont {Zen},\ and\ \citenamefont {Sorella}}]{2020NAK2}%
  \BibitemOpen
  \bibfield  {author} {\bibinfo {author} {\bibfnamefont {K.}~\bibnamefont {Nakano}}, \bibinfo {author} {\bibfnamefont {C.}~\bibnamefont {Attaccalite}}, \bibinfo {author} {\bibfnamefont {M.}~\bibnamefont {Barborini}}, \bibinfo {author} {\bibfnamefont {L.}~\bibnamefont {Capriotti}}, \bibinfo {author} {\bibfnamefont {M.}~\bibnamefont {Casula}}, \bibinfo {author} {\bibfnamefont {E.}~\bibnamefont {Coccia}}, \bibinfo {author} {\bibfnamefont {M.}~\bibnamefont {Dagrada}}, \bibinfo {author} {\bibfnamefont {C.}~\bibnamefont {Genovese}}, \bibinfo {author} {\bibfnamefont {Y.}~\bibnamefont {Luo}}, \bibinfo {author} {\bibfnamefont {G.}~\bibnamefont {Mazzola}}, \bibinfo {author} {\bibfnamefont {A.}~\bibnamefont {Zen}},\ and\ \bibinfo {author} {\bibfnamefont {S.}~\bibnamefont {Sorella}},\ }\href {https://doi.org/10.1063/5.0005037} {\bibfield  {journal} {\bibinfo  {journal} {J. Chem. Phys.}\ }\textbf {\bibinfo {volume} {152}},\ \bibinfo {pages} {204121} (\bibinfo {year} {2020})}\BibitemShut {NoStop}%
\bibitem [{Note1()}]{Note1}%
  \BibitemOpen
  \bibinfo {note} {The AGP ansatz has been generalized also to spin-polarized systems~{\cite {2003CAS}}.}\BibitemShut {Stop}%
\bibitem [{\citenamefont {Becca}\ and\ \citenamefont {Sorella}(2017)}]{2017BEC}%
  \BibitemOpen
  \bibfield  {author} {\bibinfo {author} {\bibfnamefont {F.}~\bibnamefont {Becca}}\ and\ \bibinfo {author} {\bibfnamefont {S.}~\bibnamefont {Sorella}},\ }\href@noop {} {\emph {\bibinfo {title} {{Quantum Monte Carlo approaches for correlated systems}}}}\ (\bibinfo  {publisher} {Cambridge University Press},\ \bibinfo {year} {2017})\BibitemShut {NoStop}%
\bibitem [{\citenamefont {Jensen}(2017)}]{2017JEN}%
  \BibitemOpen
  \bibfield  {author} {\bibinfo {author} {\bibfnamefont {F.}~\bibnamefont {Jensen}},\ }\href@noop {} {\emph {\bibinfo {title} {Introduction to computational chemistry}}}\ (\bibinfo  {publisher} {John wiley \& sons},\ \bibinfo {year} {2017})\BibitemShut {NoStop}%
\bibitem [{\citenamefont {Toulouse}(2018)}]{2018JUL}%
  \BibitemOpen
  \bibfield  {author} {\bibinfo {author} {\bibfnamefont {J.}~\bibnamefont {Toulouse}},\ }\href@noop {} {\bibinfo {title} {Introduction to the calculation of molecular properties by response theory}},\ \bibinfo {howpublished} {\url{https://hal.science/hal-03934866/document}} (\bibinfo {year} {2018}),\ \bibinfo {note} {[Online; accessed 19-Mar-2024]}\BibitemShut {NoStop}%
\bibitem [{\citenamefont {Goscinski}(1982)}]{1982OSV}%
  \BibitemOpen
  \bibfield  {author} {\bibinfo {author} {\bibfnamefont {O.}~\bibnamefont {Goscinski}},\ }\href@noop {} {\bibfield  {journal} {\bibinfo  {journal} {Int. J. Quantum Chem.}\ }\textbf {\bibinfo {volume} {22}},\ \bibinfo {pages} {591} (\bibinfo {year} {1982})}\BibitemShut {NoStop}%
\bibitem [{Note2()}]{Note2}%
  \BibitemOpen
  \bibinfo {note} {$\lambda _{i,j}$ is invariant under a unitary transformation of MOs only if $\lambda _{i, j} = \DOTSB \sum@ \slimits@ _{k} c_{i,k}^* c_{j,k}$ in complex cases.}\BibitemShut {Stop}%
\bibitem [{Note3()}]{Note3}%
  \BibitemOpen
  \bibinfo {note} {Only its real part is taken in complex cases~{\cite {2017BEC}}}\BibitemShut {NoStop}%
\bibitem [{\citenamefont {Pathak}\ and\ \citenamefont {Wagner}(2020)}]{2020PAT}%
  \BibitemOpen
  \bibfield  {author} {\bibinfo {author} {\bibfnamefont {S.}~\bibnamefont {Pathak}}\ and\ \bibinfo {author} {\bibfnamefont {L.~K.}\ \bibnamefont {Wagner}},\ }\href@noop {} {\bibfield  {journal} {\bibinfo  {journal} {AIP Adv.}\ }\textbf {\bibinfo {volume} {10}},\ \bibinfo {pages} {085213} (\bibinfo {year} {2020})}\BibitemShut {NoStop}%
\bibitem [{\citenamefont {Umrigar}\ and\ \citenamefont {Filippi}(2005)}]{2005UMR}%
  \BibitemOpen
  \bibfield  {author} {\bibinfo {author} {\bibfnamefont {C.~J.}\ \bibnamefont {Umrigar}}\ and\ \bibinfo {author} {\bibfnamefont {C.}~\bibnamefont {Filippi}},\ }\href {https://doi.org/10.1103/PhysRevLett.94.150201} {\bibfield  {journal} {\bibinfo  {journal} {Phys. Rev. Lett.}\ }\textbf {\bibinfo {volume} {94}},\ \bibinfo {pages} {150201} (\bibinfo {year} {2005})}\BibitemShut {NoStop}%
\bibitem [{\citenamefont {Bennett}\ \emph {et~al.}(2017)\citenamefont {Bennett}, \citenamefont {Melton}, \citenamefont {Annaberdiyev}, \citenamefont {Wang}, \citenamefont {Shulenburger},\ and\ \citenamefont {Mitas}}]{2017BEN}%
  \BibitemOpen
  \bibfield  {author} {\bibinfo {author} {\bibfnamefont {M.~C.}\ \bibnamefont {Bennett}}, \bibinfo {author} {\bibfnamefont {C.~A.}\ \bibnamefont {Melton}}, \bibinfo {author} {\bibfnamefont {A.}~\bibnamefont {Annaberdiyev}}, \bibinfo {author} {\bibfnamefont {G.}~\bibnamefont {Wang}}, \bibinfo {author} {\bibfnamefont {L.}~\bibnamefont {Shulenburger}},\ and\ \bibinfo {author} {\bibfnamefont {L.}~\bibnamefont {Mitas}},\ }\href {https://doi.org/10.1063/1.4995643} {\bibfield  {journal} {\bibinfo  {journal} {J. Chem. Phys.}\ }\textbf {\bibinfo {volume} {147}},\ \bibinfo {pages} {224106} (\bibinfo {year} {2017})}\BibitemShut {NoStop}%
\bibitem [{\citenamefont {Bennett}\ \emph {et~al.}(2018)\citenamefont {Bennett}, \citenamefont {Wang}, \citenamefont {Annaberdiyev}, \citenamefont {Melton}, \citenamefont {Shulenburger},\ and\ \citenamefont {Mitas}}]{2018BEN}%
  \BibitemOpen
  \bibfield  {author} {\bibinfo {author} {\bibfnamefont {M.~C.}\ \bibnamefont {Bennett}}, \bibinfo {author} {\bibfnamefont {G.}~\bibnamefont {Wang}}, \bibinfo {author} {\bibfnamefont {A.}~\bibnamefont {Annaberdiyev}}, \bibinfo {author} {\bibfnamefont {C.~A.}\ \bibnamefont {Melton}}, \bibinfo {author} {\bibfnamefont {L.}~\bibnamefont {Shulenburger}},\ and\ \bibinfo {author} {\bibfnamefont {L.}~\bibnamefont {Mitas}},\ }\href {https://doi.org/10.1063/1.5038135} {\bibfield  {journal} {\bibinfo  {journal} {J. Chem. Phys.}\ }\textbf {\bibinfo {volume} {149}},\ \bibinfo {pages} {104108} (\bibinfo {year} {2018})}\BibitemShut {NoStop}%
\bibitem [{\citenamefont {Annaberdiyev}\ \emph {et~al.}(2018)\citenamefont {Annaberdiyev}, \citenamefont {Wang}, \citenamefont {Melton}, \citenamefont {{Chandler Bennett}}, \citenamefont {Shulenburger},\ and\ \citenamefont {Mitas}}]{2018ANN}%
  \BibitemOpen
  \bibfield  {author} {\bibinfo {author} {\bibfnamefont {A.}~\bibnamefont {Annaberdiyev}}, \bibinfo {author} {\bibfnamefont {G.}~\bibnamefont {Wang}}, \bibinfo {author} {\bibfnamefont {C.~A.}\ \bibnamefont {Melton}}, \bibinfo {author} {\bibfnamefont {M.}~\bibnamefont {{Chandler Bennett}}}, \bibinfo {author} {\bibfnamefont {L.}~\bibnamefont {Shulenburger}},\ and\ \bibinfo {author} {\bibfnamefont {L.}~\bibnamefont {Mitas}},\ }\href {https://doi.org/10.1063/1.5040472} {\bibfield  {journal} {\bibinfo  {journal} {J. Chem. Phys.}\ }\textbf {\bibinfo {volume} {149}},\ \bibinfo {pages} {134108} (\bibinfo {year} {2018})}\BibitemShut {NoStop}%
\bibitem [{\citenamefont {Wang}\ \emph {et~al.}(2019)\citenamefont {Wang}, \citenamefont {Annaberdiyev}, \citenamefont {Melton}, \citenamefont {Bennett}, \citenamefont {Shulenburger},\ and\ \citenamefont {Mitas}}]{2019WAN}%
  \BibitemOpen
  \bibfield  {author} {\bibinfo {author} {\bibfnamefont {G.}~\bibnamefont {Wang}}, \bibinfo {author} {\bibfnamefont {A.}~\bibnamefont {Annaberdiyev}}, \bibinfo {author} {\bibfnamefont {C.~A.}\ \bibnamefont {Melton}}, \bibinfo {author} {\bibfnamefont {M.~C.}\ \bibnamefont {Bennett}}, \bibinfo {author} {\bibfnamefont {L.}~\bibnamefont {Shulenburger}},\ and\ \bibinfo {author} {\bibfnamefont {L.}~\bibnamefont {Mitas}},\ }\href {https://doi.org/10.1063/1.5121006} {\bibfield  {journal} {\bibinfo  {journal} {J. Chem. Phys.}\ }\textbf {\bibinfo {volume} {151}},\ \bibinfo {pages} {144110} (\bibinfo {year} {2019})}\BibitemShut {NoStop}%
\bibitem [{\citenamefont {Sun}\ \emph {et~al.}(2018)\citenamefont {Sun}, \citenamefont {Berkelbach}, \citenamefont {Blunt}, \citenamefont {Booth}, \citenamefont {Guo}, \citenamefont {Li}, \citenamefont {Liu}, \citenamefont {McClain}, \citenamefont {Sayfutyarova}, \citenamefont {Sharma}, \citenamefont {Wouters},\ and\ \citenamefont {Chan}}]{2018SUN}%
  \BibitemOpen
  \bibfield  {author} {\bibinfo {author} {\bibfnamefont {Q.}~\bibnamefont {Sun}}, \bibinfo {author} {\bibfnamefont {T.~C.}\ \bibnamefont {Berkelbach}}, \bibinfo {author} {\bibfnamefont {N.~S.}\ \bibnamefont {Blunt}}, \bibinfo {author} {\bibfnamefont {G.~H.}\ \bibnamefont {Booth}}, \bibinfo {author} {\bibfnamefont {S.}~\bibnamefont {Guo}}, \bibinfo {author} {\bibfnamefont {Z.}~\bibnamefont {Li}}, \bibinfo {author} {\bibfnamefont {J.}~\bibnamefont {Liu}}, \bibinfo {author} {\bibfnamefont {J.~D.}\ \bibnamefont {McClain}}, \bibinfo {author} {\bibfnamefont {E.~R.}\ \bibnamefont {Sayfutyarova}}, \bibinfo {author} {\bibfnamefont {S.}~\bibnamefont {Sharma}}, \bibinfo {author} {\bibfnamefont {S.}~\bibnamefont {Wouters}},\ and\ \bibinfo {author} {\bibfnamefont {G.~K.-L.}\ \bibnamefont {Chan}},\ }\href {https://doi.org/10.1002/wcms.1340} {\bibfield  {journal} {\bibinfo  {journal} {Wiley Interdiscip. Rev. Comput. Mol. Sci.}\ }\textbf {\bibinfo {volume} {8}},\ \bibinfo {pages} {e1340} (\bibinfo {year} {2018})}\BibitemShut
  {NoStop}%
\bibitem [{\citenamefont {Sun}\ \emph {et~al.}(2020)\citenamefont {Sun}, \citenamefont {Zhang}, \citenamefont {Banerjee}, \citenamefont {Bao}, \citenamefont {Barbry}, \citenamefont {Blunt}, \citenamefont {Bogdanov}, \citenamefont {Booth}, \citenamefont {Chen}, \citenamefont {Cui}, \citenamefont {Eriksen}, \citenamefont {Gao}, \citenamefont {Guo}, \citenamefont {Hermann}, \citenamefont {Hermes}, \citenamefont {Koh}, \citenamefont {Koval}, \citenamefont {Lehtola}, \citenamefont {Li}, \citenamefont {Liu}, \citenamefont {Mardirossian}, \citenamefont {McClain}, \citenamefont {Motta}, \citenamefont {Mussard}, \citenamefont {Pham}, \citenamefont {Pulkin}, \citenamefont {Purwanto}, \citenamefont {Robinson}, \citenamefont {Ronca}, \citenamefont {Sayfutyarova}, \citenamefont {Scheurer}, \citenamefont {Schurkus}, \citenamefont {Smith}, \citenamefont {Sun}, \citenamefont {Sun}, \citenamefont {Upadhyay}, \citenamefont {Wagner}, \citenamefont {Wang}, \citenamefont {White}, \citenamefont {Whitfield}, \citenamefont
  {Williamson}, \citenamefont {Wouters}, \citenamefont {Yang}, \citenamefont {Yu}, \citenamefont {Zhu}, \citenamefont {Berkelbach}, \citenamefont {Sharma}, \citenamefont {Sokolov},\ and\ \citenamefont {Chan}}]{2020SUN}%
  \BibitemOpen
  \bibfield  {author} {\bibinfo {author} {\bibfnamefont {Q.}~\bibnamefont {Sun}}, \bibinfo {author} {\bibfnamefont {X.}~\bibnamefont {Zhang}}, \bibinfo {author} {\bibfnamefont {S.}~\bibnamefont {Banerjee}}, \bibinfo {author} {\bibfnamefont {P.}~\bibnamefont {Bao}}, \bibinfo {author} {\bibfnamefont {M.}~\bibnamefont {Barbry}}, \bibinfo {author} {\bibfnamefont {N.~S.}\ \bibnamefont {Blunt}}, \bibinfo {author} {\bibfnamefont {N.~A.}\ \bibnamefont {Bogdanov}}, \bibinfo {author} {\bibfnamefont {G.~H.}\ \bibnamefont {Booth}}, \bibinfo {author} {\bibfnamefont {J.}~\bibnamefont {Chen}}, \bibinfo {author} {\bibfnamefont {Z.-H.}\ \bibnamefont {Cui}}, \bibinfo {author} {\bibfnamefont {J.~J.}\ \bibnamefont {Eriksen}}, \bibinfo {author} {\bibfnamefont {Y.}~\bibnamefont {Gao}}, \bibinfo {author} {\bibfnamefont {S.}~\bibnamefont {Guo}}, \bibinfo {author} {\bibfnamefont {J.}~\bibnamefont {Hermann}}, \bibinfo {author} {\bibfnamefont {M.~R.}\ \bibnamefont {Hermes}}, \bibinfo {author} {\bibfnamefont {K.}~\bibnamefont {Koh}},
  \bibinfo {author} {\bibfnamefont {P.}~\bibnamefont {Koval}}, \bibinfo {author} {\bibfnamefont {S.}~\bibnamefont {Lehtola}}, \bibinfo {author} {\bibfnamefont {Z.}~\bibnamefont {Li}}, \bibinfo {author} {\bibfnamefont {J.}~\bibnamefont {Liu}}, \bibinfo {author} {\bibfnamefont {N.}~\bibnamefont {Mardirossian}}, \bibinfo {author} {\bibfnamefont {J.~D.}\ \bibnamefont {McClain}}, \bibinfo {author} {\bibfnamefont {M.}~\bibnamefont {Motta}}, \bibinfo {author} {\bibfnamefont {B.}~\bibnamefont {Mussard}}, \bibinfo {author} {\bibfnamefont {H.~Q.}\ \bibnamefont {Pham}}, \bibinfo {author} {\bibfnamefont {A.}~\bibnamefont {Pulkin}}, \bibinfo {author} {\bibfnamefont {W.}~\bibnamefont {Purwanto}}, \bibinfo {author} {\bibfnamefont {P.~J.}\ \bibnamefont {Robinson}}, \bibinfo {author} {\bibfnamefont {E.}~\bibnamefont {Ronca}}, \bibinfo {author} {\bibfnamefont {E.~R.}\ \bibnamefont {Sayfutyarova}}, \bibinfo {author} {\bibfnamefont {M.}~\bibnamefont {Scheurer}}, \bibinfo {author} {\bibfnamefont {H.~F.}\ \bibnamefont {Schurkus}},
  \bibinfo {author} {\bibfnamefont {J.~E.~T.}\ \bibnamefont {Smith}}, \bibinfo {author} {\bibfnamefont {C.}~\bibnamefont {Sun}}, \bibinfo {author} {\bibfnamefont {S.-N.}\ \bibnamefont {Sun}}, \bibinfo {author} {\bibfnamefont {S.}~\bibnamefont {Upadhyay}}, \bibinfo {author} {\bibfnamefont {L.~K.}\ \bibnamefont {Wagner}}, \bibinfo {author} {\bibfnamefont {X.}~\bibnamefont {Wang}}, \bibinfo {author} {\bibfnamefont {A.}~\bibnamefont {White}}, \bibinfo {author} {\bibfnamefont {J.~D.}\ \bibnamefont {Whitfield}}, \bibinfo {author} {\bibfnamefont {M.~J.}\ \bibnamefont {Williamson}}, \bibinfo {author} {\bibfnamefont {S.}~\bibnamefont {Wouters}}, \bibinfo {author} {\bibfnamefont {J.}~\bibnamefont {Yang}}, \bibinfo {author} {\bibfnamefont {J.~M.}\ \bibnamefont {Yu}}, \bibinfo {author} {\bibfnamefont {T.}~\bibnamefont {Zhu}}, \bibinfo {author} {\bibfnamefont {T.~C.}\ \bibnamefont {Berkelbach}}, \bibinfo {author} {\bibfnamefont {S.}~\bibnamefont {Sharma}}, \bibinfo {author} {\bibfnamefont {A.~Y.}\ \bibnamefont
  {Sokolov}},\ and\ \bibinfo {author} {\bibfnamefont {G.~K.-L.}\ \bibnamefont {Chan}},\ }\href {https://doi.org/10.1063/5.0006074} {\bibfield  {journal} {\bibinfo  {journal} {J. Chem. Phys.}\ }\textbf {\bibinfo {volume} {153}},\ \bibinfo {pages} {024109} (\bibinfo {year} {2020})}\BibitemShut {NoStop}%
\bibitem [{\citenamefont {Perdew}\ and\ \citenamefont {Zunger}(1981)}]{1981PER}%
  \BibitemOpen
  \bibfield  {author} {\bibinfo {author} {\bibfnamefont {J.~P.}\ \bibnamefont {Perdew}}\ and\ \bibinfo {author} {\bibfnamefont {A.}~\bibnamefont {Zunger}},\ }\href {https://doi.org/10.1103/PhysRevB.23.5048} {\bibfield  {journal} {\bibinfo  {journal} {Phys. Rev. B}\ }\textbf {\bibinfo {volume} {23}},\ \bibinfo {pages} {5048} (\bibinfo {year} {1981})}\BibitemShut {NoStop}%
\bibitem [{\citenamefont {Nakano}\ \emph {et~al.}(2023)\citenamefont {Nakano}, \citenamefont {Kohulák}, \citenamefont {Raghav}, \citenamefont {Casula},\ and\ \citenamefont {Sorella}}]{2023NAK}%
  \BibitemOpen
  \bibfield  {author} {\bibinfo {author} {\bibfnamefont {K.}~\bibnamefont {Nakano}}, \bibinfo {author} {\bibfnamefont {O.}~\bibnamefont {Kohulák}}, \bibinfo {author} {\bibfnamefont {A.}~\bibnamefont {Raghav}}, \bibinfo {author} {\bibfnamefont {M.}~\bibnamefont {Casula}},\ and\ \bibinfo {author} {\bibfnamefont {S.}~\bibnamefont {Sorella}},\ }\href {https://doi.org/10.1063/5.0179003} {\bibfield  {journal} {\bibinfo  {journal} {J. Chem. Phys.}\ }\textbf {\bibinfo {volume} {159}},\ \bibinfo {pages} {224801} (\bibinfo {year} {2023})}\BibitemShut {NoStop}%
\bibitem [{\citenamefont {Posenitskiy}\ \emph {et~al.}(2023)\citenamefont {Posenitskiy}, \citenamefont {Chilkuri}, \citenamefont {Ammar}, \citenamefont {Hapka}, \citenamefont {Pernal}, \citenamefont {Shinde}, \citenamefont {Landinez~Borda}, \citenamefont {Filippi}, \citenamefont {Nakano}, \citenamefont {Kohulák}, \citenamefont {Sorella}, \citenamefont {de~Oliveira~Castro}, \citenamefont {Jalby}, \citenamefont {Ríos}, \citenamefont {Alavi},\ and\ \citenamefont {Scemama}}]{2023POS}%
  \BibitemOpen
  \bibfield  {author} {\bibinfo {author} {\bibfnamefont {E.}~\bibnamefont {Posenitskiy}}, \bibinfo {author} {\bibfnamefont {V.~G.}\ \bibnamefont {Chilkuri}}, \bibinfo {author} {\bibfnamefont {A.}~\bibnamefont {Ammar}}, \bibinfo {author} {\bibfnamefont {M.}~\bibnamefont {Hapka}}, \bibinfo {author} {\bibfnamefont {K.}~\bibnamefont {Pernal}}, \bibinfo {author} {\bibfnamefont {R.}~\bibnamefont {Shinde}}, \bibinfo {author} {\bibfnamefont {E.~J.}\ \bibnamefont {Landinez~Borda}}, \bibinfo {author} {\bibfnamefont {C.}~\bibnamefont {Filippi}}, \bibinfo {author} {\bibfnamefont {K.}~\bibnamefont {Nakano}}, \bibinfo {author} {\bibfnamefont {O.}~\bibnamefont {Kohulák}}, \bibinfo {author} {\bibfnamefont {S.}~\bibnamefont {Sorella}}, \bibinfo {author} {\bibfnamefont {P.}~\bibnamefont {de~Oliveira~Castro}}, \bibinfo {author} {\bibfnamefont {W.}~\bibnamefont {Jalby}}, \bibinfo {author} {\bibfnamefont {P.~L.}\ \bibnamefont {Ríos}}, \bibinfo {author} {\bibfnamefont {A.}~\bibnamefont {Alavi}},\ and\ \bibinfo {author}
  {\bibfnamefont {A.}~\bibnamefont {Scemama}},\ }\href {https://doi.org/10.1063/5.0148161} {\bibfield  {journal} {\bibinfo  {journal} {J. Chem. Phys.}\ }\textbf {\bibinfo {volume} {158}},\ \bibinfo {pages} {174801} (\bibinfo {year} {2023})}\BibitemShut {NoStop}%
\bibitem [{Note4()}]{Note4}%
  \BibitemOpen
  \bibinfo {note} {See Ref.{~\cite {2020NAK2}} for the detail of the Jastrow factor implementation in the {\protect \emph {ab initio}} QMC package, \protect \textsc {TurboRVB}, used in this study.}\BibitemShut {Stop}%
\bibitem [{\citenamefont {Huber}(2013)}]{2013HUB}%
  \BibitemOpen
  \bibfield  {author} {\bibinfo {author} {\bibfnamefont {K.-P.}\ \bibnamefont {Huber}},\ }\href {https://doi.org/10.1119/1.1932852} {\emph {\bibinfo {title} {{Molecular spectra and molecular structure: IV. Constants of diatomic molecules}}}}\ (\bibinfo  {publisher} {Springer Science \& Business Media},\ \bibinfo {year} {2013})\BibitemShut {NoStop}%
\bibitem [{\citenamefont {Nakano}\ \emph {et~al.}(2021)\citenamefont {Nakano}, \citenamefont {Morresi}, \citenamefont {Casula}, \citenamefont {Maezono},\ and\ \citenamefont {Sorella}}]{2021NAK1}%
  \BibitemOpen
  \bibfield  {author} {\bibinfo {author} {\bibfnamefont {K.}~\bibnamefont {Nakano}}, \bibinfo {author} {\bibfnamefont {T.}~\bibnamefont {Morresi}}, \bibinfo {author} {\bibfnamefont {M.}~\bibnamefont {Casula}}, \bibinfo {author} {\bibfnamefont {R.}~\bibnamefont {Maezono}},\ and\ \bibinfo {author} {\bibfnamefont {S.}~\bibnamefont {Sorella}},\ }\href {https://doi.org/10.1103/PhysRevB.103.L121110} {\bibfield  {journal} {\bibinfo  {journal} {Phys. Rev. B}\ }\textbf {\bibinfo {volume} {103}},\ \bibinfo {pages} {L121110} (\bibinfo {year} {2021})}\BibitemShut {NoStop}%
\bibitem [{\citenamefont {Vinet}\ \emph {et~al.}(1987)\citenamefont {Vinet}, \citenamefont {Smith}, \citenamefont {Ferrante},\ and\ \citenamefont {Rose}}]{1987VIN}%
  \BibitemOpen
  \bibfield  {author} {\bibinfo {author} {\bibfnamefont {P.}~\bibnamefont {Vinet}}, \bibinfo {author} {\bibfnamefont {J.~R.}\ \bibnamefont {Smith}}, \bibinfo {author} {\bibfnamefont {J.}~\bibnamefont {Ferrante}},\ and\ \bibinfo {author} {\bibfnamefont {J.~H.}\ \bibnamefont {Rose}},\ }\href {https://doi.org/10.1103/PhysRevB.35.1945} {\bibfield  {journal} {\bibinfo  {journal} {Phys. Rev. B}\ }\textbf {\bibinfo {volume} {35}},\ \bibinfo {pages} {1945} (\bibinfo {year} {1987})}\BibitemShut {NoStop}%
\bibitem [{Note5()}]{Note5}%
  \BibitemOpen
  \bibinfo {note} {In the variance estimation, covariance effects are neglected. Nevertheless, in the actual calculations, the full statistical error was estimated by the Jackknife method~{\cite {2017BEC}}.}\BibitemShut {Stop}%
\bibitem [{\citenamefont {Momma}\ and\ \citenamefont {Izumi}(2011)}]{2011MOM}%
  \BibitemOpen
  \bibfield  {author} {\bibinfo {author} {\bibfnamefont {K.}~\bibnamefont {Momma}}\ and\ \bibinfo {author} {\bibfnamefont {F.}~\bibnamefont {Izumi}},\ }\href@noop {} {\bibfield  {journal} {\bibinfo  {journal} {J. Appl. Crystallogr.}\ }\textbf {\bibinfo {volume} {44}},\ \bibinfo {pages} {1272} (\bibinfo {year} {2011})}\BibitemShut {NoStop}%
\end{thebibliography}%

\end{document}